\documentclass[11pt]{article}
\pdfoutput=1
\usepackage{a4wide}
\usepackage{amsmath,amssymb,amsfonts}
\usepackage{comment}
\usepackage{epsfig,verbatim,graphics}
\usepackage{hyperref}
\usepackage{slashed}
\usepackage{caption}
\usepackage{subcaption}
\newcommand{\mathsym}[1]{{}}

\newcommand{\eref}[1]{(\ref{#1})}

\renewcommand\({\left(}
\renewcommand\){\right)}
\renewcommand\[{\left[}
\renewcommand\]{\right]}

\newcommand{\dd}{{\rm d}}
\declareslashed{}{{^-}}{.1}{0.1}{\rm d}

\newcommand{\e}{{\rm e}}

\newcommand\eps{\epsilon}
\newcommand\mpl{M_{\rm P}}

\def\be{\begin{equation}}
\def\ee{\end{equation}}
\def\ba{\begin{eqnarray}}
\def\ea{\end{eqnarray}}

\def\L{\mathcal{L}}
\def\G{\mathcal{G}}
\def\O{\mathcal{O}}
\def\A{\mathcal{A}}

\def\M{\mathcal{M}}

\def\U{\mathcal{U}}

\def\del{\nabla}
\def\nn{\nonumber}
\def\({\left(}
\def\){\right)}

\def\Tr{{\rm Tr}}

\usepackage{tikz}
\usetikzlibrary{arrows,shapes}
\usetikzlibrary{trees}
\usetikzlibrary{matrix,arrows} 				
\usetikzlibrary{positioning}				
\usetikzlibrary{calc,through}				
\usetikzlibrary{decorations.pathreplacing}  
\usepackage{pgffor}							

\usetikzlibrary{decorations.pathmorphing}	
\usetikzlibrary{decorations.markings}
\tikzset{
	vector/.style={decorate, decoration={snake}, draw},
	provector/.style={decorate, decoration={snake,amplitude=2.5pt}, draw},
	antivector/.style={decorate, decoration={snake,amplitude=-2.5pt}, draw},
	fermion/.style={draw=black, postaction={decorate},
		decoration={markings,mark=at position .55 with {\arrow[draw=black]{>}}}},
	fermionbar/.style={draw=black, postaction={decorate},
		decoration={markings,mark=at position .55 with {\arrow[draw=black]{<}}}},
	fermionnoarrow/.style={draw=black},
	gluon/.style={decorate, draw=black,
		decoration={coil,amplitude=4pt, segment length=5pt}},
	scalar/.style={dashed,draw=black, postaction={decorate},
		decoration={markings,mark=at position .55 with {\arrow[draw=black]{>}}}},
	scalarbar/.style={dashed,draw=black, postaction={decorate},
		decoration={markings,mark=at position .55 with {\arrow[draw=black]{<}}}},
	scalarnoarrow/.style={dashed,draw=black},
	electron/.style={draw=black, postaction={decorate},
		decoration={markings,mark=at position .55 with {\arrow[draw=black]{>}}}},
	bigvector/.style={decorate, decoration={snake,amplitude=4pt}, draw},
}

\newcommand{\roughly}[1]{\mathrel{\raise.3ex\hbox{$#1$\kern-0.85em
			\lower1ex\hbox{$\sim$}}}}

\begin{document}

\begin{titlepage}
\begin{center}

\hfill Nikhef 2017-054\\

\vskip 1.5cm

{\LARGE \bf Unitarity and predictiveness in new Higgs inflation}

\vskip 1cm

\renewcommand*{\thefootnote}{\fnsymbol{footnote}}
\setcounter{footnote}{0}

{\bf
    Jacopo\ Fumagalli$^{1}$,
    Sander Mooij$^{2}$ 
and
    Marieke Postma$^1$
}

\renewcommand*{\thefootnote}{\number{footnote}}
\setcounter{footnote}{0}

\vskip 25pt

{\em $^1$ \hskip -.1truecm
Nikhef, \\Science Park 105, \\1098 XG Amsterdam, The Netherlands
}


\vskip 20pt
{\em $^2$ \hskip -.1truecm
Institute of Physics, Laboratory of Particle Physics and Cosmology, \'Ecole Polytechnique F\'ed\'erale de Lausanne, CH-1015 Lausanne, Switzerland 
}

\end{center}
\vskip 0.5cm
\begin{abstract}In new Higgs inflation  the Higgs kinetic terms are non-minimally coupled to the Einstein tensor, allowing the Higgs field to play the role of the inflaton. The new interaction is non-renormalizable, and the model only describes physics below some cutoff scale. Even if the unknown UV physics does not affect the tree level inflaton potential significantly, it may still enter at loop level and modify the running of the Standard Model (SM) parameters. This is analogous to what happens in the original model for Higgs inflation.  A key difference, though, is that in new Higgs inflation the inflationary predictions are sensitive to this running. Thus the boundary conditions at the EW scale as well as the unknown UV completion may leave a signature on the inflationary parameters. However, this dependence can be evaded if the kinetic terms of the SM  fermions and gauge fields are non-minimally coupled to gravity as well.
Our approach to determine the model's UV dependence and the connection between low and high scale physics can be used in any particle physics model of inflation.
\end{abstract}

\end{titlepage}

\newpage
\setcounter{page}{1}%
\tableofcontents



\section{Introduction}

Models in which the Standard Model (SM) Higgs field plays the role of the inflaton are attractive.  First, the Higgs field is the only scalar we know to exist, and thus this approach can be considered minimal as no new field has to be  assumed.  Second, it offers the possibility to test Higgs physics at high energies relevant for inflation, and to link the data from the Cosmic Microwave Background to what is measured in colliders at low energies. Given the very different energy scales involved, the running of the SM parameters has to be taken into account for a sensible comparison. 

Although it does not give a valid period of inflation, for illustrative purposes consider first inflation in the SM with the quartic Higgs potential $V(\phi)=\lambda \phi^4$.  The SM is renormalizable, and the running of the couplings can be computed straightforwardly.  Using the re\-nor\-ma\-li\-za\-tion group (RG) improved potential for the inflationary analysis,  the prediction for the scalar spectral index $n_s$ is
\begin{equation}\label{gencorrection}
n_s=n_{s0}\left(1+\kappa \frac{\beta_{\lambda}}{\lambda}\right),
\end{equation} 
where $n_{s0}$ stands for the observable computed at tree level, $\beta_\lambda$ is the beta function for the quartic coupling $\lambda$, and $\kappa =1$ for the quartic potential. Likewise, the tensor-to-scalar ratio will get a correction due to the running of $\lambda$.  The Higgs field has superplanckian field values during inflation, and consequently all SM fermions and gauge bosons are very heavy during inflation and can be integrated out \cite{BasteroGil:2010vq}. With only the Higgs field itself (and the Goldstone bosons) in the spectrum, the beta function during the inflationary epoch is $\beta_{\lambda}\propto\lambda^2/(8\pi^2)$ and the corrections are always small for perturbative values of the coupling. In fact, for quartic inflation, $\lambda\sim O(10^{-14})$ is required to fit the CMB amplitude \cite{Linde:1983gd} --- which is why it does not work in the SM --- and the correction is completely negligible.

For the Higgs to be the inflaton new interactions beyond those already present in the SM are needed.  In the original Higgs inflation (HI) proposal \cite{fakir,salopek,bezrukov1} a non-minimal coupling between the Higgs and the Ricci tensor is introduced, in new Higgs inflation (NHI) \cite{Germani:2010gm} the Higgs kinetic terms are non-minimally coupled to the Einstein tensor, and in Higgs G-inflation \cite{Kobayashi:2010cm,Kamada:2010qe} and running kinetic Higgs inflation \cite{Nakayama:2010sk,Nakayama:2014koa} a non-minimal kinetic coupling for the Higgs is considered. These are all examples of the general class of Horndeski type of interactions \cite{Horndeski:1974wa,Kamada:2012se}. The fact that these new interactions are non-renormalizable changes the story described above in two important ways.  First, for consistency of the theory new physics below the Planck scale is needed,\footnote{In general, ``new physics'' can either be new fields and interactions, or that the theory becomes strongly coupled \cite{ufuk} and the perturbative analysis breaks down.} which opens the possibility that the renormalization group equations are modified by the unknown UV completion.  Second, thanks to the non-minimal coupling, during inflation the fermions and gauge bosons might be light enough to remain in the spectrum, and the beta function will be of the form $\beta_{\lambda} \propto g^4$ with $g$ a Yukawa or gauge coupling.  Nothing prevents the ratio $\beta_{\lambda}/\lambda$ from being sizeable, and the running corrections to the observables of the form \eref{gencorrection} can be large.  The two effects combined introduce a UV sensitivity into the model \cite{cliffnew}.  In this paper we discuss the UV (in)sensitivity of NHI, and contrast it with the results for HI discussed in \cite{Fumagalli,Fumagalli:2016sof}.

It is worth clarifying that the UV sensitivity we are interested in here is not the common one related to unknown Planck scale physics. This triggers the $\eta$-problem and, more general, the sensitivity to Planck scale suppressed higher dimensional operators that correct the inflaton potential. We treat this kind of UV contributions as part of the definition of the model assuming these corrections are sufficiently small,  either due to some symmetry principle (e.g. an approximate shift symmetry),  a one-time fine-tuning, or because of the specific nature of the Planck scale physics.

One way to see the need for new physics is to consider the tree level scattering amplitudes, and to see at what energy scale unitarity is lost.  This gives the unitarity bound of the theory, which in these set-ups is field dependent \cite{bezrukov4,Ferrara:2010in}.  The field dependent cutoff, even if always above the typical energy scales along the field domain, might be close to the inflationary scale in part of the field regime. This happens generically in the mid field regime, indicating that especially in this region the new physics needed to restore unitarity can be important. 
It should also be noted that in both HI and NHI the unitarity bound in the mid field regime is lower than the typical potential energy scale during inflation. As mentioned above, we work under the assumption that the UV completion does not affect the inflationary potential at tree-level significantly, but it is clearly not guaranteed \textit{a priori} that what solves the unitarity problem (i.e. what lifts the unitarity bound to the Planck scale over the whole field range) will not affect the inflationary regime.\footnote{In \cite{Giudice:2010ka,Barbon:2015fla} HI UV completions were discussed but in these models the Higgs does not play the role of the inflaton.}

Although the non-minimal couplings in HI and NHI are non-renormalizable, a re\-nor\-ma\-li\-zable effective field theory (EFT) can be constructed in both the small and large field regime.
 This means that when the Lagrangian is expanded in a small parameter, all loop divergencies can be absorbed in counterterms order by order in the expansion parameter \cite{bezrukov4, George:2015nza}.
The counterterms are different in both regimes though, and in the mid field region threshold corrections are needed to patch the EFTs together. Higher order operators are needed to absorb the divergences in the mid field regime, breaking the connection between the low/high scale parameters \cite{bezrukov4,shap,critical1}.

The threshold corrections can be parametrized by an infinite tower of higher order ope\-ra\-tors suppressed by the unitarity cutoff \cite{cliffnew,critical1,Fumagalli}, which may become relevant in the mid field regime.  These operators give corrections to the RGE equations, which depending on the Wilson coefficients can be large in the transition region. The net effect will be a ``kick'' in the values of the running couplings.  As a result the values of the couplings during inflation will depend both on the boundary conditions at the electroweak scale (the measured values), and on the UV completion via this kick.

The story so far is equally valid for both HI and NHI (although the details, such as the explicit values of the cutoff and the form of the RGEs differ).  The main difference between the two scenarios is the UV sensitivity of the inflationary observables. In HI, the running corrections to the spectral index and the tensor-to-scalar ratio are suppressed to leading order in slow roll through a cancellation of different effects \cite{Fumagalli}.  This particular feature is shared by the more general class of Cosmological Attractors \cite{Fumagalli:2016sof} of which HI is a particular case. Thus the spectral index is given by \eref{gencorrection} with $\kappa \approx 0$.  Thus whatever the kick, whatever the effect of the UV completion on the running,\footnote{This holds as long as inflation takes place in the ``universal'' regime deep inside the large field region; if instead the potential is tuned and inflation takes place in the mid field regime
close to an extremum generated by quantum corrections the model is UV sensitive as well \cite{Fumagalli,Fumagalli:2016sof,Enckell:2016xse, Bezrukov:2017dyv}.} the predictions for $n_s$ and $r$ are given by the tree-level results, and thus robust.  In NHI inflation on the other hand, as we will show in this paper, this is not the case;  in this set-up $\kappa = \O(1)$ and the model is UV sensitive. This conclusion can be avoided if the fermions and gauge bosons are coupled non-minimally to gravity as well. In that case, they are very light during inflation and effectively decouple. Just as for the example of the quartic potential discussed above, the inflationary predictions are not affected by the uncertainty in the running of the couplings. 
 
This paper is organized as follows. In section \ref{sec2} we give a quick review of the new Higgs inflation proposal. In section \ref{rg dependence} we compute the inflationary parameters taking into account the running of the couplings. In \ref{sec4} we discuss the unitarity of the model, complementing the analysis done in \cite{germaniU}, in order to check whether the typical energy scale is close to the tree level cutoff of the theory. In \ref{s:renormalizability} the (non-) renormalizability of the model is discussed with special emphasis on the threshold corrections required to consistently connect the two asymptotic regimes.  We end in \ref{conclusions} with some conclusive remarks and we compare the results to those of the original Higgs inflation model. Appendix \ref{A:improved} provides additional details on the RG improved action.

\section{New Higgs inflation: a quick review} \label{sec2}

In this section we give the action for new Higgs inflation (NHI), show how its analysis can be simplified after performing a disformal transformation,  identify the different field regimes, and study the inflationary dynamics.

\subsection{The action}

In NHI the Higgs kinetic terms are non-minimally coupled to gravity \cite{Germani:2010gm}. We will also consider the possibility that the kinetic terms of the fermions and gauge bosons contain a non-minimal coupling to gravity \cite{Germani:2011cv,disformal}.  To assure second order equations of motion for both gravity and matter, the new couplings should be to divergenceless tensors constructed from the
Riemann tensor. In four dimensions there is only the
Einstein tensor $G_{\mu\nu} = R_{\mu\nu} -(R/2)g_{\mu\nu}$, and the
double dual Riemann
tensor\footnote{$\epsilon^{abcd} = \frac{1}{\sqrt{-g}} \e^{abcd}$,
  with $\epsilon^{abcd} $ the Levi-Civita tensor and $\e^{abcd}$ the
  completely antisymmetric symbol.} \cite{dual1,dual2}
$\G_{\alpha \beta\gamma\delta} =* R_{\alpha \beta\gamma\delta} *=
\frac14 \eps_{\alpha \beta \mu \nu} R^{\mu\nu\rho \sigma}\eps_{\rho
  \sigma\gamma\delta } $,
which satisfy $\del_\mu G^{\mu\nu} =0$ and
$\del_\mu \G^{\mu\nu\rho \sigma} =0$.

The action is
\be
S= \int \dd^4x \sqrt{-\bar g}
\[ \frac12 \mpl^2 \bar R + \L_{\rm Higgs}+\L_{\rm fermion} +\L_{\rm gauge}\].
\label{Ltot}
\ee
For reasons that become clear in a moment we denote the FLRW metric by an overbar $\bar{g}_{\mu\nu} = {\rm diag}(-1,\bar a^2, \bar a^2 ,\bar a^2)$. The Higgs, fermion and gauge
Lagrangian contain the SM terms plus a non-minimal coupling to
gravity via the kinetic term.  Explicitly \cite{disformal}
\begin{align}
\L_{\rm Higgs} &= -\( \bar
g^{\mu\nu} - \alpha_\phi \frac{\bar G^{\mu\nu}}{M^2} \) D_\mu \Phi^\dagger D_\nu \Phi -
V  \nn \\
\L_{\rm gauge} &= - \sum \frac14 \(\bar g^{\alpha\mu} \bar g^{\beta
  \nu} - \alpha_A 
\frac{3\bar \G^{\mu\nu \alpha\beta}}{M^2}
\)F^a_{\alpha\beta}F^a_{\mu\nu}
\nn \\
\L_{\rm fermion} &= -
\sum_i \(\bar g^{\alpha \beta} - \alpha_\psi \frac{\bar G^{\alpha\beta}}{M^2} \) \bar \psi_i
  i\gamma_\alpha D_\beta \psi_i
- (y_d \bar Q_L \Phi d_R + y_u \bar Q_L \tilde \Phi u_R   + y_e \bar L_L \Phi e_R + {\rm h.c.} ).
\label{ActionNH}
\end{align}
with $\tilde \Phi =(i\sigma^2)\Phi^*$. The summation in the gauge Lagrangian runs over the SM gauge groups.  The SM fermion fields are $\psi_i= \{Q_L,u_R,d_R,E_L,e_R\}$: the left-handed doublet, right-handed up and down quark, left-handed lepton, right-handed electron respectively (we suppressed family indices).  The couplings $\alpha_i$ can be Higgs field dependent: $\alpha_i=\alpha_i(\Phi)$. NHI assumes a constant Higgs coupling, which can always be set to unity by redefining the mass scale $M$.  From now on we set $\alpha_\phi=1$.  In the original NHI scenario this is the only non-minimal coupling and $\alpha_A=\alpha_\psi=0$ \cite{Germani:2010gm}.  However, one can consider the more general possibility $\alpha_i =\O(1) \(\frac{\sqrt{V}}{M \mpl}\)^n$ with $i=\psi,A$.

\subsection{Disformal transformation}

The Higgs-gravity sector can be brought in (approximate) standard form:  an Einstein-Hilbert term plus a scalar field Lagrangian.  Consider a disformal transformation of the metric \cite{disformal}
\be
g_{\alpha\beta} = \bar g_{\alpha\beta}+\varepsilon_{\alpha\beta}
= \bar g_{\alpha\beta}-2\frac{D_\alpha \Phi^\dagger
  D_\beta \Phi}{\M^4},
\ee
where we introduced the scale
\be
\M^2 = M \mpl .
\label{Lambda}
\ee
In the small derivative expansion regime we have $\varepsilon \ll 1$.  To describe the evolution of the classical field $\langle |\Phi|^2 \rangle = \phi(t)^2/2$ the expansion is certainly valid.  Indeed, evaluated on the background \cite{Ema:2015oaa}
\be
\varepsilon^0_0 = \frac{\dot \phi^2}{\M^4}
= \frac{2(3\bar H^2 \mpl^2 - V)}{9\bar H^2 \mpl^2+ \M^4}< \frac23,
\label{eps0}
\ee
where in the second step we have used the Friedmann equations.  During inflation the potential dominates the energy density and the numerator $3\bar H^2 \mpl^2 - V \ll 3\bar H^2 \mpl^2 $ is small.  In this regime $ \varepsilon^0_0 \sim {\eps} $ the slow roll parameter. At small field values $\bar H^2 \mpl^2 \ll \M^4$ and $\varepsilon^0_0 \ll 1$ as well.  The maximum $\varepsilon^0_0 \approx \frac23$ is reached at the end of inflation/onset of preheating.

To leading order the action after the disformal transformation is
\begin{align}
\sqrt{-\bar g}\(\bar R+ 2\bar G^{\mu\nu} \frac{D_\mu \Phi^\dagger
D_\nu \Phi}{\M^4} \) &= \sqrt{- g}R +\O(\varepsilon), \nn \\
\sqrt{-\bar g} &= \sqrt{-g}\(1 + \frac{D_\alpha \Phi^\dagger
D^\alpha \Phi}{\M^4} +\O(\varepsilon^2)\), \nn \\
\bar g^{\alpha \beta} &=  g^{\alpha \beta} -2 \frac{D^\alpha \Phi^\dagger
D^\beta \Phi}{\M^4} +\O(\varepsilon^2).
\end{align}
The gravity-Higgs sector transforms into standard Einstein gravity
plus the Higgs Lagrangian; the effect of the original Higgs-gravity coupling is now transferred to non-minimal kinetic terms
\begin{align}
S_{\rm EH + Higgs}  & =  \int \dd^4x \sqrt{-g} \[ \frac12 \mpl^2 R - 
\(1+ \frac{V}{\M^4}\) D^\mu \Phi^\dagger
D_\mu \Phi -V +\O(\varepsilon^2)\] .
\label{Lhiggs} 
\end{align}
To leading order the gauge and fermion Lagrangians are invariant, and we can replace $\bar g_{\mu \nu} \to g_{\mu\nu} $ in $\L_{\rm gauge}$ and $\L_{\rm fermion}$ in \eref{ActionNH}.  On the background the double-dual Riemann tensor is $ {\G^{0i}}_{0i} =-H^2$ and $ { G^{ij}}_{ij} =-{\ddot{a}}/{a} = -H^2(1-\eps)$, and the Einstein tensor is $G^\mu_\nu =-3H^2 {\rm diag}(1,1-\frac23 \eps )$, with $ \eps = -{\dot H}/{H^2}$.  During inflation $\eps \sim \varepsilon_0^0$ and the $\varepsilon$-terms can be dropped at leading order.  After inflation, taking the scale factor as a power law $a \propto t^n$ with $n<1$, one finds $\eps =1/n$.   Lorentz symmetry is broken spontaneously by the background by order one effects in the fermion and gauge kinetic terms.  If we are interested in order of magnitude estimates, we can ignore these effects.  The gauge and fermion Lagrangians then take the form
\begin{align}
S_{\rm gauge}
&=
 -\frac14 \int \dd^4 x \sqrt{- g}\sum \(g^{\alpha\mu}  g^{\beta \nu} -
  \alpha_A \frac{ 3\G^{\mu\nu\alpha\beta}}{M^2}\)
F^a_{\mu\nu} F^a_{\alpha\beta} + \O(\varepsilon)
\nn \\
&\sim -\frac14 \int \dd^4 x \sqrt{- g}\sum \(1 +
  \alpha_A \frac{V}{\M^4}\)
F^a_{\mu\nu} F^{a, \mu\nu} ,
\label{Lgauge} 
\end{align}
and 
\begin{align}
\L_{\rm fermion} & = 
-\sum_i \(g^{\alpha \beta} - \alpha_\psi \frac{ G^{\alpha\beta}}{M^2} \) \bar \psi_i
  i\gamma_\alpha D_\beta \psi_i
- (y_t \bar Q_L \tilde \Phi t_R+{\rm h.c.} ) +\O(\varepsilon)
\nn\\
&\sim- \sum_i \(1+ \alpha_\psi \frac{ V}{\M^4} \) \bar \psi_i
  i\gamma^\mu D_\mu \psi_i
- ( y_t \bar Q_L \tilde \Phi t_R  +{\rm h.c.} ) ,
\label{Lfermion}
\end{align}
where for concreteness we have focussed on the top quark Lagrangian, but all Yukawa interactions have the same structure.

\subsection{Summary and notation}

To summarize, and set the notation used in the next sections, we work with the NHI Lagrangian in the Einstein frame given in (\ref{Lhiggs},~\ref{Lgauge},~\ref{Lfermion}) 
\begin{align}
\label{L}
\L =
-\gamma(\Phi) |D_\mu \Phi|^2- V(\Phi)-\sum_a \frac14 k^2(\Phi)(F^a_{\mu\nu})^2 
+ \sum_i q^2(\Phi)\bar \psi_i (i\slashed{D}) \psi_i - \frac{y_t}{\sqrt{2}} \bar Q_L \tilde \Phi t_R +{\rm h.c.}
\end{align}
where the sum is over all SM gauge groups; we only included the top quark and the sum in the fermion kinetic terms is over $\psi_i =\{Q_L,t_R\}$.  The functions $\gamma,k^2,q^2$ are given explicitly in \eref{functions}.  In unitary gauge the Higgs doublet is parametrized
\be
\Phi =
\frac1{\sqrt{2}} \( \begin{array}{c} 0  \\
  \phi_r\end{array} \)  ,
\label{Hfields}
\ee
with the real Higgs field $\phi_r = \phi + \delta \phi$ split in a background field $\phi$ plus fluctuations $\delta \phi$.  We denote the canonically normalized Higgs field by $h_r = h + \delta h$, with $h$ the classical background.

The dynamics of the system is very different in the small and large
field regimes, where the correction to the Higgs kinetic term is not
important respectively dominates.  Defining 
\be
\delta \equiv \frac{V}{\M^4}
\label{delta}
\ee
we distinguish between the small field ($\delta \ll 1)$ and large field ($\delta \gg 1)$ regime.
The boundary between the large and small field regime is at $\delta =1$
for field values
\be
\phi_{\rm eq} = \sqrt{2} \frac{\M}{\lambda^{1/4}}.
\label{phi_eq}
\ee 
The non-minimal Higgs, gauge and fermion field space metrics in \eref{L} can then be written
\begin{align}\label{functions}
 \gamma &= \(1+ \delta \), &
q^2 &= \(1+ \alpha_F \delta\) ,&
k^2 &= \(1+ \alpha_A \delta\) .
\end{align}
We parameterize the non-minimal gauge boson and fermion couplings as
\be
\alpha_i  = \alpha_{0i} \delta^{n_i/2}
\label{alpha}
\ee
for $i=A,F$.

\subsection{Inflation}

The Higgs-gravity action
for the classical background is
\be
\L=\sqrt{-g} \left[\frac{1}{2}\mpl^2 R
- \frac12\gamma (\phi) (\partial_\mu \phi)^2
-\frac{\lambda}{4} \phi^4 \],
\label{L_inf}
\ee
where we have neglected the Higgs mass term during inflation.
In the large field regime, the canonically normalized field $h$ is
\be
\partial_\mu h \approx \sqrt{\delta}
\partial_\mu \phi \quad \rightarrow \quad h = 
\frac{\sqrt{\lambda}\phi^3}{6\M^2} +C
\label{dh}.
\ee
The integration constant $C$ can be fixed by matching it to the small field solution at the boundary region $h(\phi_{\rm eq}) = \phi_{\rm eq}$.  At large field values this constant can be neglected, and we ignore it from now on.
In terms of the canonical field the Lagrangian then is
\be
\L=\sqrt{-g} \left[\frac{1}{2}\mpl^2 R -\frac{1}{2}\partial_\mu
  h \partial^\mu h
-\frac{\tilde{\lambda}}{4} \mpl^4  \(\frac{h}{\mpl}\)^{4/3} \right],
\label{chaL}
\ee
with
\be 
\tilde{\lambda} =6^{4/3} \lambda^{1/3} 
\(\frac{\M}{\mpl}\)^{8/3}.
\label{lambdat}
\ee
For large field values the theory is nothing but old-fashioned canonical chaotic inflation, albeit with a rather unconventional exponent of $4/3$ in the potential.

\section{Renormalization group dependence of inflationary predictions}\label{rg dependence}

In this section we derive NHI's predictions for $n_s$ and $r$. The results at tree level (first reported in \cite{Germani:2010ux,disformal}) are just the standard ones for chaotic inflation. We include corrections due to the running of both $\lambda$ and $\M$, which is a new result. We work with the one-loop RG improved effective potential $V = V(h,\lambda(\mu), \M(\mu))$ with running couplings.  The calculation can be done either in terms of the field $\phi$ or using the canonical field $h$, the results are the same.  Since it is rather subtle to show the equivalence, we give more details in Appendix \ref{A:improved}.  Here we choose to work with the canonical field as it makes the integral determining the number of efolds below trivial.

As the renormalization scale we choose\footnote{This choice minimizes the leading logs in the Coleman-Weinberg potential (see section \ref{Sm spectrum}, in both case A and B). We could have chosen $\mu  = y_t \phi$, which leads to a $y_t$-dependence.  However, now the
  matching conditions (boundary conditions for the running couplings) at the EW
  scale are different. In the end one obtains the same result for the physical observables, which indeed should  be independent of the renormalization scale.}
\be
\mu \sim m_t \sim \phi=\mpl^{2/3}\left(\frac{\tilde\lambda}{\lambda}\right)^{1/4}h^{1/3}.
\label{mu}
\ee
To take the effect of the running couplings into account we note that the derivative of the coupling is proportional to the beta function
\be
\frac{d \lambda(\mu)}{d h}= \frac{d
  \lambda(\mu)}{d t}
\frac{d t}{d h}  \equiv \beta_\lambda \frac{d t}{d h} ,\qquad
\frac{d\beta_{\lambda}}{dt}\equiv\beta_\lambda',
\ee
and likewise for $\tilde \lambda$ and $M$.  Here $t = \ln(\mu/m_t^{\rm EW})$ is the renormalization time, and
\be
\frac{dt}{dh}=\frac{1}{3h}+\left(\frac{\beta_{\tilde\lambda}}{4\tilde\lambda}-\frac{\beta_{\lambda}}{4\lambda}\right)\frac{dt}{dh}\implies \frac{dt}{dh}=\frac{1}{3h}\left(1+\frac{\beta_{\lambda}}{4\lambda}-\frac{\beta_{\tilde\lambda}}{4\tilde\lambda}\right)^{-1}
\ee
where we used the explict form of the renormalization scale \eref{mu}.

The potential slow roll parameters are
\be 
\eps_V \equiv \frac{\mpl^2}{2} \left(\frac{V_h}{V}\right)^2
= \frac{8\mpl^2}{9 h^2}\left(1+2\frac{\beta_{\tilde\lambda}}{4\tilde\lambda}\right)
+\mathcal{O}\left(\frac{\beta_X'}{4X},\frac{\beta_X}{4X}\frac{\beta_Y}{4Y}\right),
\label{slow1}
\ee
and
\be
\eta_V \equiv \mpl^2 \frac{V_{hh}}{V}\simeq \frac{4\mpl^2}{9 h^2}\left(1+5\frac{\beta_{\tilde{\lambda}}}{4\tilde{\lambda}}\right)+\mathcal{O}\left(\frac{\beta_X'}{4X},\frac{\beta_X}{4X}\frac{\beta_Y}{4Y}\right).
\label{slow2}
\ee
Here we used the approximation 
 \be\label{approx}
 \frac{\beta_{X}^{(n)}}{4X},\frac{\beta_{X}^2}{(4 X)^2} \ll 
 \frac{\beta_{X}}{4X}
 \ee
for $X,Y =\lambda,\tilde\lambda,M$, and $\beta_X^{(n)}$ the nth derivative with respect to the renormalization time $t$. This inequality is satisfied for the SM beta function at high scales, but should be checked explicitly in NHI, as the new interactions will affect the running at high scales. As follows from \eref{observables} derived below, the running corrections become order one precisely at the boundary of the region of validity of the approximation \eref{approx}.  This analytical approximation, therefore, is only valid for small, albeit not negligible, corrections. To analyze larger corrections one would need to turn to numerics in order to properly calculate the observables.

For the inflationary observables we need the slow roll parameters evaluated $N_\star$
efolds before the end of inflation:
\be
N_{\star}
\simeq \frac{1}{\mpl}\int_{h_{\rm end}}^{h_*} \dd h {\frac{1
	}{\sqrt{2\eps_V}}} \simeq
\frac{3}{4\mpl^2}\int_{h_{\rm end}}^{h_*} \dd h\,h\,D\, ,
\qquad {\rm with} \;\;
D=\(1-\frac{\beta_{\tilde\lambda}}{4\tilde\lambda}\).
\label{Nexpand}
\ee
%
%
The approximation \eref{approx} allows us to consistently consider $D\simeq D_{\star}$ almost constant over the integration domain. Thus, the integral becomes trivial and we can easily solve for the number of efolds,
\be
N_{\star}\simeq \frac{3}{8 \mpl^2}h^2_{\star} D_{\star}.
\ee
The slow roll parameters in \eref{slow1} and \eref{slow2}, evaluated $N_\star$ efolds before the end of inflation are then
\be
\eps_{V\star} \simeq \frac1{3N_\star} \(1 +\frac{\beta_{\tilde\lambda}}{4\tilde\lambda}\)_{\star},\quad
\eta_{V\star} \simeq \frac1{6N_\star} \(1 +4\frac{\beta_{\tilde\lambda}}{4\tilde\lambda}\)_{\star}.
\label{sr_star}
\ee
Finally, to leading order in the $1/N_\star$ expansion the observables are:
\be
n_s-1 = 2\eta_{V\star}-6\eps_{V\star} \simeq -\frac5{3N_\star}\(1 +\frac{2}{5}\frac{\beta_{\tilde\lambda}}{4\tilde\lambda}\)_\star,
\quad
r=16 \eps_{V\star} \simeq \frac{16}{3N_\star} \(1 +\frac{\beta_{\tilde\lambda}}{4\tilde\lambda}\)_\star
\label{observables}
\ee
up to
$\mathcal{O}\left(\frac1{N_\star^2},\frac{\beta_X'}{X},\frac{\beta_X}{X}\frac{\beta_Y}{Y}\right)$ corrections. It is interesting to note that, within the approximation \eref{approx}, the explicit dependence on the RG flow is only through the running of $\tilde{\lambda}$, i.e. the self coupling of the canonical Higgs field in the large field regime. This is not obvious a priori since we have used a renormalization scale of the form \eref{mu}.

The running corrections can become large if the ratio $\beta_{\tilde \lambda}/(4{\tilde \lambda})$ is order one during inflation. We will estimate the typical size of $\beta_{\tilde \lambda}$  in section \ref{s:renormalizability}. We stress that the above result is markedly different from that in the original Higgs inflation scenario.  In HI, the RG corrections to $n_s$ and $r$ disappear at first order in the $1/N_\star$ expansion due to a cancellation between the running dependence of the slow roll parameters and of the number of efolds \cite{Fumagalli,Fumagalli:2016sof}. In NHI, in contrast, such a cancellation does not take place. Therefore, the influence of the RG flow (and consequently, the influence of the theory's UV completion) on the inflationary predictions in  NHI is parametrically larger than in  HI.

To end the inflationary analysis, we give here the explicit values for the scales involved.  Taking $N_* =60$ at tree level the spectral index and tensor-to-scalar ratio \eref{observables} are $n_s \simeq0.97$ and $r\simeq 0.09$.  The field value during and at the end of inflation is $\phi_\star^6 = ( 96 N_* \M^4 \mpl^2/\lambda)=(3 N_\star) \phi_{\rm end}^6$. The power spectrum fixes the free parameter $\M$ in the theory via ${V}/({\mpl^4 \eps})|_\star = (0.027)^4$ \cite{Planck}, and we find
\begin{align}
M & \simeq 1.5 \times 10^{-8} \mpl \lambda^{-1/4}, &
\M &\simeq 1.2 \times 10^{-4} \mpl \lambda^{-1/8},\nn \\
\delta_* &\simeq 1.3 \times 10^{7} \sqrt{\lambda},&
\phi_* &\simeq 1.0 \times 10^{-2} \mpl \lambda^{-1/4}.
\label{numbersf}
\end{align}
Since $\delta_{\rm{end}}\simeq 4\times10^5\sqrt{\lambda}$, the end of inflation is well inside the large field regime. 

%
%


\section{Unitarity bound}\label{sec4}

The semiclassical approximation is valid if typical energy scales are below the scale at which tree level unitarity breaks down.  Because of the non-renormalizable interactions in NHI the model becomes ill-defined at large scales. Expanding the Lagrangian around $\phi=0$, the higher order interactions are suppressed by the cutoff scale $\sim \M$.  Likewise, one can expand the Lagrangian around large field values $\phi$, and read off the typical cutoff scale from the non-renormalizable interaction.  The cutoff in NHI is field dependent, and different in the asymptotic regimes.  A more systematic approach to determine the cutoff scale for the validity of the theory is to calculate the scale at which tree-level unitarity is lost, obtained from scattering amplitudes/cross sections for specific interactions. In this section we compute the 2-to-2 Higgs/Goldstone boson scattering; for non-minimally coupled fermions and (transverse) gauge bosons, it is in addition interesting to look at their scattering rates. We compare to the results for $2h \to nh$ scattering, with $h$ the (canonically normalized) Higgs field, obtained in \cite{germaniU} and find order one agreement.

We first recap the general approach for extracting the unitarity bounds of the Standard Model EFT in chiral
representation \cite{contino}, which was applied to Higgs inflation in \cite{cliffnew}.

\subsection{Chiral Standard Model with non-minimal gauge/fermion sector}

In the chiral approach the Higgs doublet is parametrized
\be
H = \frac{\phi_r}{\sqrt{2}} \e^{-i \vec \sigma . \vec \chi/F_0}
= \frac{\phi_r}{\sqrt{2}} \, \U,
\label{U}
\ee
with $\phi_r = \phi +\delta \phi $ the Higgs field split in its background value and fluctuations, and $\vec \chi$ the Goldstone bosons.  $F_0$ is a normalization constant to give the Goldstone bosons canonical kinetic terms.  The kinetic terms for the Higgs are non-minimal, and it is convenient to rewrite the Lagrangian in terms of the canonical Higgs denoted by $h_r(\phi_r) =h +\delta h$. If the top quark and gauge bosons are non-minimally coupled, their kinetic terms are of the form\footnote{Since we are only interested in the order of magnitude of the unitarity bound, we neglect Lorentz violating effects and use the approximations (\ref{Lgauge},\ref{Lfermion}).  We assume universal $\alpha_A$ couplings for all gauge bosons, and universal $\alpha_F$  couplings for all fermions.  The results are easy to generalize.}
\begin{align}
\L &\supset 
-\sum \frac14 k^2(h_r)  F_{\mu \nu}^2 +
q^2(h_r)  (\bar Q_L i\slashed{ D} Q_L + \bar t_R \slashed{ D} t_R).
\label{L_uni}
\end{align}
%
We rescale
\begin{align}
A_\mu =  \frac{\tilde A_\mu}{k_0}, \quad 
Q_L = \frac{\tilde Q_L}{q_0} , \quad
t_R = \frac{\tilde t_R}{q_0} , \quad
g = k_0  \tilde g ,\quad
y_t = q_0^2 \tilde y_t,
\label{Atilde}
\end{align}
with $g$ and $y_t$ the gauge and top Yukawa coupling, and we have used the notation $f_0 \equiv f(h)$, i.e. the function evaluated at its background value. With this rescaling the top and gauge boson fields have canonical quadratic kinetic terms, and the gauge and Yukawa interactions are still in standard form as $g A_\mu = \tilde g \tilde A_\mu$ and $y_t \bar Q_L d_R = \tilde y_t \tilde{\bar{Q}}_L \tilde d_R$.  This assures the Goldstone boson equivalence theorem holds as usual.
The Lagrangian is then
\begin{align}
\L &= -\frac12 (\partial h_r)^2 -\frac14 F^2(h_r) \Tr \[(\tilde D_\mu \U)^\dagger
(\tilde D^\mu \U) \] 
-\sum \frac14 \frac{k^2(h_r)}{k_0^2} \tilde F_{\mu \nu}^2 +
\frac{q^2(h_r)}{q_0^2}   (\tilde{\bar{ Q}}_L \slashed{\tilde  D}  \tilde Q_L 
+ \tilde{\bar{t}}_R \slashed{\tilde  D} \tilde t_R)
\nn \\& \hspace{0.4cm}
-V(h_r)- \frac1{\sqrt{2}}  \( y_t \tilde{\bar{Q}}_L (i\sigma^2)
\U^* \tilde t_R\) Y(h_r) + {\rm h.c.} .
\label{L_uni2}
\end{align}
For notational convenience we will drop the tildes, but in the rest of the section we will always work with the Lagrangian \eref{L_uni2}.  The amplitude for $\chi^+ \chi^- \to \bar \psi \psi$ mediated by the Yukawa\footnote{In NHI $Y= \phi_R$.} interaction does  not give a strong bound for NHI, and we will not consider it explicitly. Next, expand\footnote{In the SM one has $F = Y = h_r =\phi_r$, and the expansion is formulated in terms of $\delta h$. }  
\begin{align}
F^2 &= F_0^2 \(1 +2a \frac{\delta h}{F_0} +b\frac{(\delta h)^2}{F_0^2} \) , &
\frac{k^2}{k_0^2} & = \(1+ \frac{k_1}{F_0} \delta h + ...\) ,
& \frac{q^2}{q_0^2}&= \(1+ \frac{q_1}{F_0} \delta h + ...\)  .
\end{align}
The coefficients $a,b,k_1,q_1$ are dimensionless. In the SM $F_0 =Y_0= h=v$, and $a=b=1$ and $k_1 =q_1 =0$. In terms of the non-canonical $\phi$-field --- related to the canonical field $h$ via $\frac12\gamma (\partial \phi)^2 = \frac12(\partial h)^2$ --- we can extract the coefficients in the expansion as follows:
\begin{align}
a 
&=  \frac1{2F} \frac1{\sqrt{\gamma}}\frac{\partial F^2}{\partial \phi_r} 
  \bigg|_{\phi},&
b& = \frac1{2} \frac1{\sqrt{\gamma}} 
\frac{\partial}{\partial \phi_r} \( \frac1{\sqrt{\gamma}}\frac{\partial
  F^2}{\partial \phi_r} \)   \bigg|_{\phi} ,&
k_1&= \frac{F_0}{ k_0^2} \frac1{\sqrt{\gamma}}\frac{\partial
  k^2}{\partial \phi_r}  
\bigg|_{\phi},&
q_1&= \frac{F_0}{q_0^2} \frac1{\sqrt{\gamma}}\frac{\partial
  q^2}{\partial \phi_r}  
\bigg|_{\phi},
\label{coeff}
\end{align}
all evaluated on the background at $\phi_r = \phi$.

The amplitude for $2 \to 2$ scattering of Goldstone bosons into Goldstone bosons and  Higgs fields can then be expressed \cite{contino}
\begin{align}
\A_1(\chi^+\chi^- \to \chi^+\chi^-) &= \frac{s+t}{F_0^2} (1-a^2) +
\O\(\frac{m_h^2}{E^2}\), \nn \\
\A_2(\chi^+\chi^- \to \delta h \, \delta h) &= \frac{s}{F_0^2} (b-a^2) +
\O\(\frac{m_h^2}{E^2}\)  .
\label{A12}
\end{align}
%
In the Standard Model all amplitudes vanish up to $\O(m_h^2/E^2)$, but in NHI the amplitudes are non-zero because of the new interactions from the non-minimal Higgs-gravity coupling.
In addition there are extra diagrams because of the non-minimal gauge- and fermion-gravity couplings, mediated by the $k_1$ and $q_1$ interactions respectively.  The $k_1$-term in the Lagrangian gives a $h (\partial A)^2$-vertex. For the scattering of longitudinal gauge bosons this interaction provides amplitudes that do not grow with energy, but as $\O(M_W^4/E^4)$. The terms proportional to $\O(s/M^2_W)$ cancel  in agreement with the Goldstone equivalence theorem (the Goldstone and thus the longitudinal gauge boson interactions do not depend on the non-minimal gauge-gravity coupling). The $k_1$-interaction still gives a growing contribution for the transverse gauge boson scattering (via the diagram with Higgs exchange).  The $q_1$-term generates a $h\bar\psi \psi$ interaction; this provides an extra contribution to the $\chi \chi \to \psi \psi$ process.  The additional amplitudes are given by
\begin{align}
\A_A (AA \to AA) &\sim \frac{s}{F_0^2}k_1^2 , \nn \\
\A_F(\chi^+\chi^- \to \bar \psi \psi) 
&\sim \frac{\sqrt{s}}{F_0} aq_1 .
\end{align}

The $2\to 2$ scattering amplitudes are of the form $\A(s,t,u) = \A(s,\theta)$; in the relativistic limit $\vec p \gg m$ the Mandelstam variables are $t =-( s/2)(1-\cos \theta)$ and $u=- (s/2)(1+\cos \theta)$.  To determine the unitarity bound we project the amplitude onto partial waves
\be
a_l = \frac1{32\pi} \int_{-1}^{-1} \dd \cos \theta \, \A(s,\theta)
P_l(\cos \theta),
\ee
with $P_l$ the Legendre polynomial $P_0(x) =1,\, P_1(x) = x$.  The
s-wave unitarity bound is $|a_0| < \pi/2$.\footnote{Also $|a_0| < 1$ and
  $|a_0| < 1/2$ are used sometimes; all are acceptable as estimates, the spread
  can be seen as theoretical uncertainty \cite{contino}. In \cite{germaniU} instead the bound $\sigma < 4\pi/s$ is used. This gives a cutoff that is lower, and thus a bound that is stronger, by about a factor of 2 for NHI.
  }
Consider first the bound for Goldstone boson scattering, for which we find
\be
a_0 = \frac1{32\pi}\frac{|1-a^2|}{F_0^2} s < \frac{\pi}{2}
\quad
\Rightarrow
\quad
\sqrt{s} < \Lambda_1 = 4\pi \frac{F_0}{\sqrt{|1-a^2|}},
\ee
which gives the unitarity  bound on the center of mass energy $\sqrt{s} = E_{\rm
  cm}$.  The bound is a factor $\sqrt{2}$ stronger if $\chi^+\chi^- \to \chi^0 \chi^0$ is included.  To find the unitarity bound for the other processes, we proceed in the same way. The result is
\begin{align}
\Lambda_1 &= 4 \pi\frac{F_0}{\sqrt{|1-a^2|}},&
\Lambda_2 &=2 \sqrt{2} \pi \frac{F_0}{\sqrt{|b-a^2|}},&
\Lambda_A &\sim 2\sqrt{2}\pi \frac{F_0}{|k_1|},&
\Lambda_F &\sim 8\pi^2 \frac{F_0}{|a q_1|}.
\label{Lambda12}
\end{align}
The SM is renormalizable, and indeed all bounds diverge for the SM values of the coefficients. The SM without the Higgs boson has $a=b=0$, and the above expressions give the unitarity cutoff of the Fermi theory.

In  \cite{germaniU}  the unitarity bound derived from $2\, \delta h \to n\, \delta h$
scattering was derived (with $n\geq 3$) using the criterion on the cross section
$\sigma < 4\pi/s$. This gives
\be
\Lambda_m \sim |V_m| F_m^{\frac{1}{2(m-2)}},
\quad
F_m = 2^{4m-10} \pi^{2m -6} (m-3)! (m-4)!,
\quad
V_m = \(\frac{\dd^m V}{\dd h^m}\)^{\frac{1}{4-m}},
\label{cutoff_m}
\ee
with $m=n+2$.

%

\subsection{New Higgs inflation}

The Lagrangian of NHI \eref{L} is of the form of the chiral SM with
\begin{align}\label{functions2}
 \gamma &= \(1+ \delta \), &
 F^2 &= \phi_r^2 \(1+ \delta\),&
q^2 &= \(1+ \alpha_F \delta\) ,&
k^2 &= \(1+ \alpha_A \delta\) ,
\end{align}
and $\delta = V/\M^4$, see \eref{delta}, and the non-minimal gauge boson and fermion couplings $\alpha_i$ are given in \eref{alpha}.
The coefficients in the expansion of the Lagrangian \eref{coeff} are then 
\begin{align}
a&= \frac{\(1+ 3\delta\)}{\(1+ \delta\)} ,&
b &=\frac{\(1+ 14 \delta
+ \(3\delta \)^2 \)}{\(1+ \delta\)^2},&
k_1& = \frac{2(2+n_A) \alpha_{A} \delta}{1+ \alpha_{A} \delta},&
q_1& =  \frac{2(2+n_F) \alpha_{F} \delta}{1+ \alpha_{F} \delta},
\end{align}
with $\delta$ evaluated on the background (for notational convenience we dropped the subscript `0').  Consider first the bounds from the Higgs sector, that are always present in NHI.  For $\delta \gg 1$, the coefficients $a,b$ approach an $\O(1)$ constant, and the unitarity cutoff for Higgs and Goldstone scales as $\Lambda \propto F_0 \propto \sqrt{\delta}$.  This estimate is actually too naive for Higgs scattering, as the leading term in the denominator in $\Lambda_2$ cancels in the large field regime, and consequently the cutoff grows faster: $\Lambda_2 \propto \delta$.  In the small field regime $a,b\to 1$ approach the SM values, and the cutoff scales as $\Lambda \propto 1/\sqrt{\delta}$.  Thus both in the limit of small and large field values, the bound increases, with a minimum at the midfield region $\delta \sim 1$. Numerically, the bound from scattering into Higgses is slightly stronger, which we give here explicitly
\begin{align}
\Lambda_2 &= \pi \phi \frac{(1+\delta)^{3/2}}{\sqrt{\delta}}
= \pi \phi
\left\{
\frac{1}{\sqrt{\delta}}, \, \delta
\right\},
\end{align}
with the right most expression the approximation in the small and large field limit respectively.  The minimum of the cutoff is in the midfield regime for $\delta =1/5$, and is given by
\begin{align}
\Lambda_{2,{\rm min}} &  \approx 6 \phi_{\rm eq} \approx 6 \times 10^{-3} \mpl \(\frac{10^{-2}}{\lambda}\)^{3/8} , 
\end{align}
with $\phi_{\rm eq}$ given in \eref{phi_eq}.  It is interesting to compare this minimum with the potential at the same point, and with the potential during inflation
\begin{align}
\frac{V^{1/4}(\phi_{\rm min})}{\Lambda_{2,{\rm min}} }
 \approx 2.4 \times 10^{-2} \(\frac{\lambda}{10^{-2}}\)^{1/4},
\qquad \qquad
\frac{V_\star^{1/4}}{\Lambda_{2,{\rm min}} }
\approx 1.2
\(\frac{\lambda}{10^{-2}}\)^{3/8}.
\label{UV}
\end{align}
where we used \eref{numbersf}.  We confirm the conclusion reached in \cite{germaniU} that $V^{1/4} < \Lambda$ for all field values, and an EFT can be constructed for energies below the cutoff.  However, it should be noted that the inflationary energy is of the same order as the minimum value of the cutoff, unless $\lambda \ll 10^{-2}$ tuned to small values during inflation.  Thus, as in the original HI proposal, it is not guaranteed \textit{a priori} that what lifts the unitarity bound to the Planck scale over the whole field range will not affect the inflationary regime already at tree level. We assume this is not the case and the UV completion  only affects the running.

\begin{figure}
	\begin{subfigure}{1\textwidth}
		\centering
		\includegraphics[width = 0.49
	\linewidth, bb = 0 4 359 225]{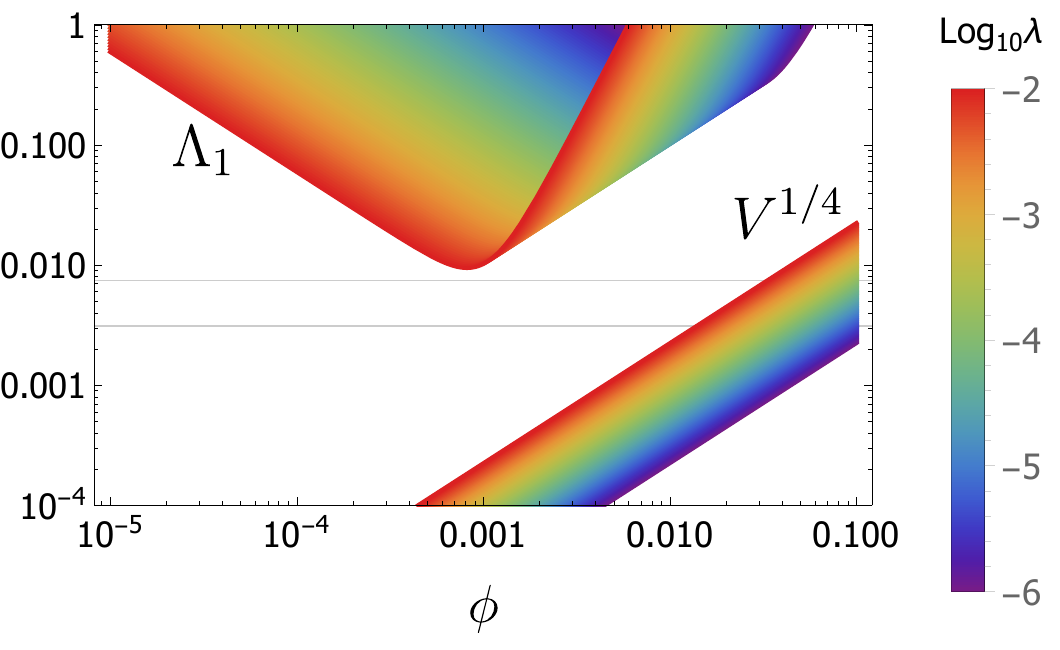}
		\includegraphics[width = 0.49\linewidth,bb = 1 3 359 222]{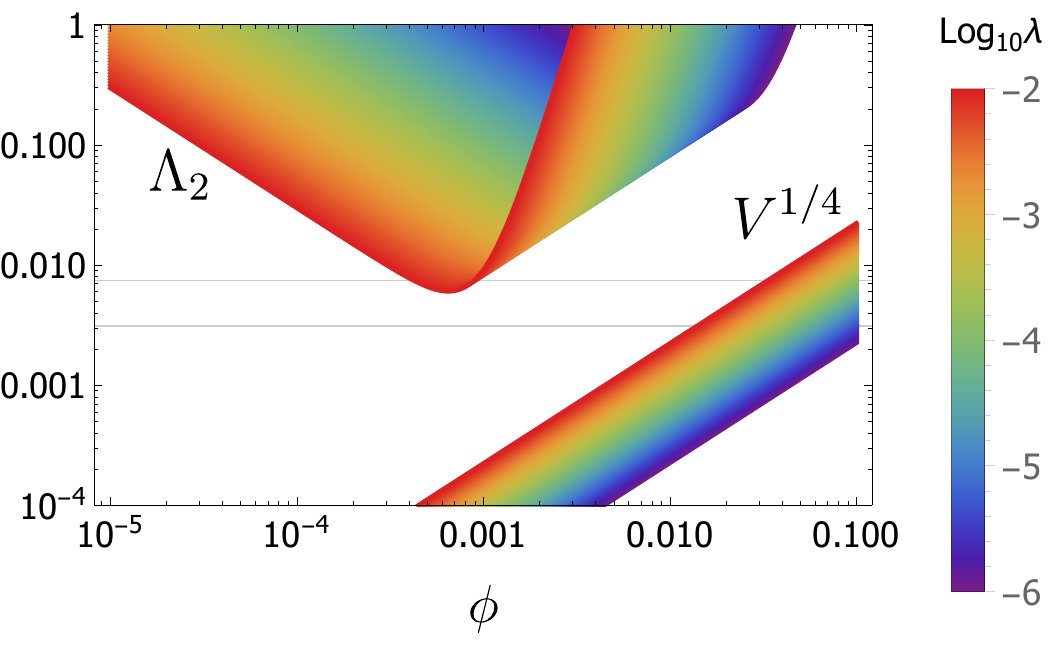}
		\vspace{.2cm}
\caption{
		Cutoff from $2 \to 2$ scattering of Goldstone bosons into Goldstone bosons ($\Lambda_1$ in the left plot) and  Higgs fields  ($\Lambda_2$ in the right plot).}
		\label{fig1}
	\end{subfigure} \\
	\begin{subfigure}{1\textwidth}
		\centering
		\includegraphics[width = 0.49\linewidth, bb = 1 4 358 239]{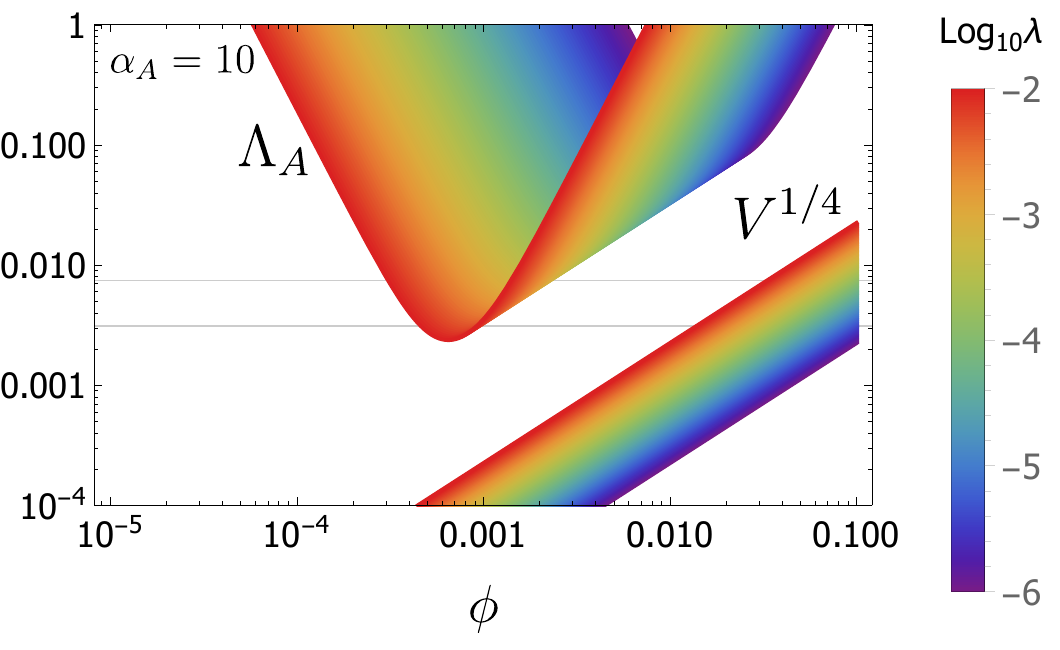}
		\includegraphics[width = 0.49\linewidth,bb = 1 4 358 239]{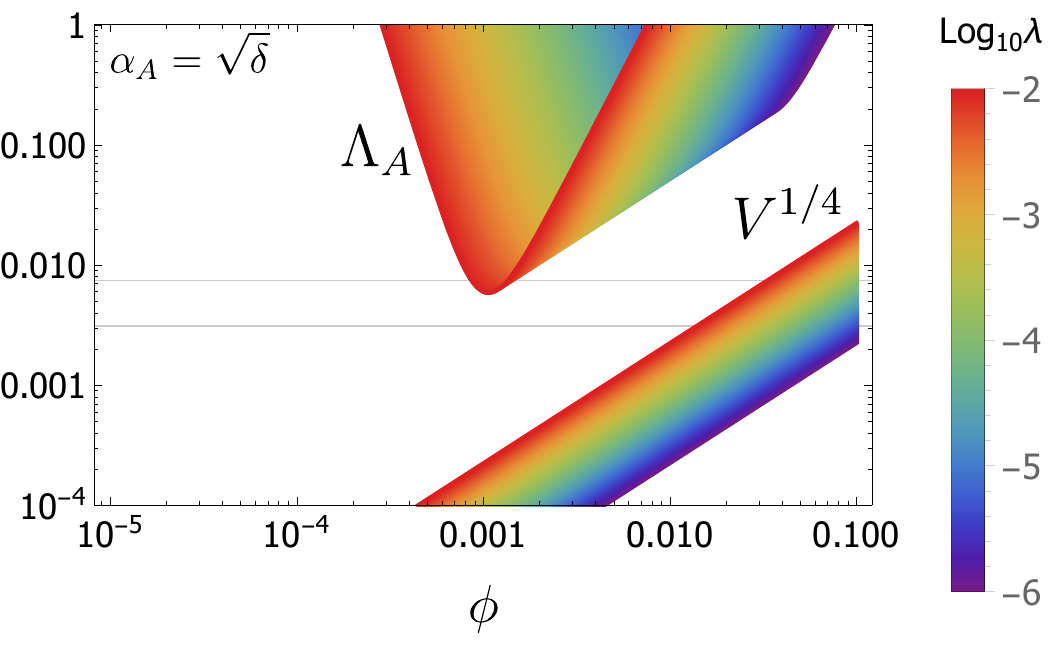}
		\vspace{-.2cm}
			\vspace{.2cm}
		\caption{Cutoff from $2 \to 2$ scattering of non-minimally coupled gauge bosons, for different non-minimal coupling $\alpha_A=10$ (left plot) and $\alpha_A = \sqrt{\delta}$ (right plot).}
		\label{fig2} 
	\end{subfigure}\\
	\begin{subfigure}{1\textwidth}
\centering
		\includegraphics[width = 0.49\linewidth, bb= 0 1 356 239]{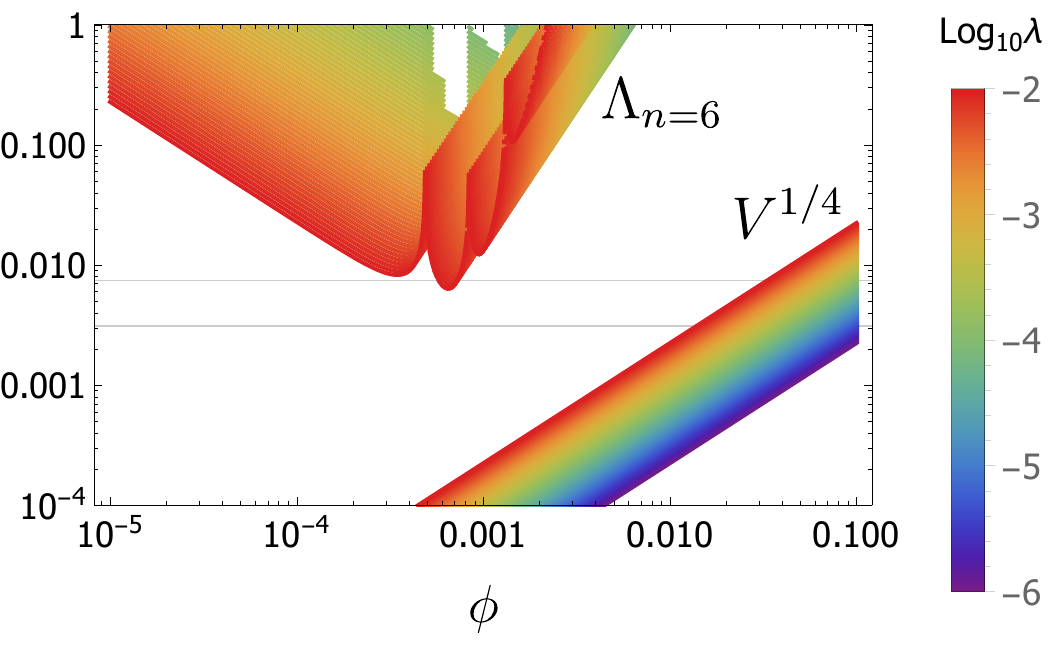}
		\includegraphics[width = 0.49\linewidth,bb= 1 1 356 239]{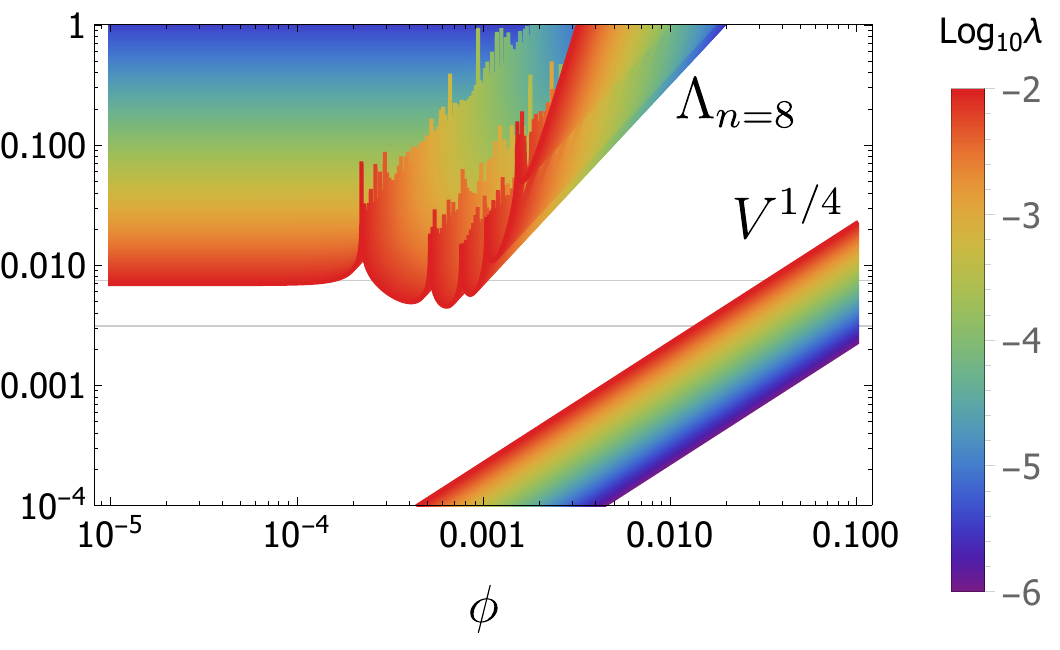}
			\vspace{.2cm}
\caption{Cutoff from $2 \to n$ Higgs scattering, with $n=6$ (left plot)  and $n= 8$ (right plot).}
\label{fig:cutoff} 
\end{subfigure}
	\vspace{.5cm}
\caption{Unitarity cutoff $\Lambda_i$ for different interactions as a function of the field $\phi$ in Planck units; also shown is $V^{1/4}$. 
		In all figures
		horizontal gridlines correspond to $\{V_{\rm end}^{1/4}, V^{1/4}_\star\}$ which are independent of $\lambda$. In the plots $\lambda$ varies between $10^{-2} $ and $10^{-6}$.}
	\label{gig11}
\end{figure}

The results  for the unitarity cutoff are shown in Fig. ~\ref{gig11} the quartic coupling $\lambda$ in the range $10^{-2}$ to $10^{-6}$.  Note that $\lambda(\phi)$ is a running coupling whose value is scale dependent.
Figure \ref{fig1} shows $\Lambda_1$ and $\Lambda_2$; for comparison also the energy scale set by the potential  $V^{1/4}$ is shown. As can easily be seen, lower values of $\lambda$ raises the cutoff and lowers the potential, with the net result that the tension between the cutoff and the inflationary scale is relieved.  
If the gauge bosons and/or fermions are non-minimally coupled there are additional bounds \eref{Lambda12}. In the small field limit $k_1 \simeq 2(1+n_A) \alpha_{0A} \delta$ and the cutoff $\Lambda_A$ increases rapidly for small $\delta$ as $\Lambda_A \propto \phi /(\alpha_{0A} \delta)$. In the large field limit $k_1 \simeq 2(1+n_A)$ and the bound is $\Lambda_A \propto \phi\sqrt{\delta}/(2(1+n_A) )$. The cutoff is minimal in the midfield regime $\delta \sim 1$, and scales as $\Lambda_{A,{\rm min}} \propto \phi_{\rm eq}/((4+2n_A)\alpha_{0A})$. That is, the larger $n_A$ and the larger $\alpha_{0A}$ the smaller the unitarity cutoff, and in this limit the bound can become stronger than the one from the Higgs sector. The behavior for the cutoff $\Lambda_F$ from fermion scattering is similar, except that it contains an extra factor of $a$ in the denominator, which suppresses the bound for similar $n_i$ and $\alpha_{0i}$ values.  For example, for $n_i=1$ and $\alpha_{0i}=1$, we find $\Lambda_{A,{\rm min}} \approx 6.3 \phi_{\rm eq}$ and $\Lambda_{F ,{\rm min}} \approx 260 \phi_{\rm eq}$.  The results are shown in Fig.~\ref{fig2} which plots the bounds from gauge boson scattering for a constant and field dependent non-minimal coupling $\alpha_{A}$.

Finally, there are the bounds from $hh \to nh$ scattering \eref{cutoff_m}. These bounds are plotted for two representative values $n=6$ and $n=6$ in Fig.~\ref{fig:cutoff}.  The minimum cutoff is of the same order as $\Lambda_{2,{\rm min}}$. However, this many-body amplitude gives a lower cutoff in the small field regime than the other processes. 

The power spectrum fixes $\tilde \lambda$ defined in \eref{lambdat}. This means that for smaller Higgs coupling during inflation $\lambda_\star$, the scale $\M$ where the non-minimal couplings enter becomes larger \eref{numbersf}. This has the effect, as can be also seen from the plots, that the hierarchy between the cutoff scale and the scale set by the potential increases for small coupling.  In fact, for $\lambda =10^{-6}$ the cutoff  is well above the  inflationary scale $\Lambda_{\rm min} \gg V_\star^{1/4}$, suggesting that the effects of the unknown UV completion are small. This is, however, not necessarily the case.  First of all, as we discuss in the next section, the light degrees of freedom in the EFT differ in the small field and large field regime, and consequently there is a jump in counterterms between the two regimes. This jump can be seen as the effect of threshold corrections, which therefore cannot be arbitrarily small (especially if the jump is proportional to the gauge or Yukawa coupling, and not only to the quartic coupling --- which depends on the possible non-minimal couplings of the gauge and fermion fields).

Secondly, the loop corrections may grow for small coupling (this also depends on the possible non-minimal couplings of the gauge and fermion fields, as discussed in the next section), and if they become large they may alter the shape of the potential significantly.  This is indeed what happens in conventional Higgs inflation, where it was found that for $\lambda_\star \lesssim 10^{-5}$ no inflationary solution with $N_\star \gtrsim 60$ is possible in the flat plateau regime of the potential \cite{kyle,Fumagalli}. However, for small coupling it is still possible to have inflation, but near an extremum of the potential where the tree level and one-loop corrections are tuned; this is the idea of ``critical Higgs inflation" \cite{Hamada:2014iga,Bezrukov:2014bra,Hamada:2014wna}. Since it is a fine-tuned set-up, the UV sensitivity is large \cite{Fumagalli,Enckell:2016xse,Bezrukov:2017dyv}. The conditions for a critical new Higgs inflation regime will most likely be analogous to those of HI, where two conditions have to be imposed on the beta functions in order to tune both the first and the second derivative of the potential to zero. A derivation of the RG equations for NHI, and an in-depth analysis of the inflationary predictions is left for future work.


\section{Renormalizability of non-renormalizable models}
\label{s:renormalizability}

In the previous section we have (re)-established that the tree-level unitarity cutoff is below the Planck scale in the small and mid field regime, as $\Lambda \sim \phi_{\rm eq} \sim \M/\lambda^{1/4}$.  Even though this cutoff is always above the typical scale set by the potential, the inflationary predictions can nevertheless be sensitive to the UV completion, as we discuss in this section.  The existence of a field dependent cutoff which is sub-Planckian in the small/mid field regime is of course no surprise, as the troubles are caused by the introduction of non-renormalizabe interactions in the NHI model that are suppressed by the scale $\M$.

 Even though the model is non-renormalizable it may still be renormalizable in the EFT sense in the asymptotic regions, by which we mean that all divergencies can be absorbed order by order in the counterterms already present in the model.\footnote{To fully check this we should calculate the full set of (one-loop) corrections; if the set-up is EFT renormalizable we can then derive the full set of RG equations in the asymptotic regimes.  This is left for future work.}  In this section we concentrate on the one-loop Coleman-Weinberg potential, and check whether the divergencies can be absorbed in counterterms of the tree-level potential.  This will give us information about the size of $\beta_{\tilde \lambda}/{\tilde \lambda}$, indicating whether the running corrections to the inflationary observables \eref{observables} can be important.  Moreover, as we will see, in the large field regime some SM fields decouple. This implies that the counterterms are different in the small and large field regime, and threshold corrections are needed in the midfield regime to patch the EFTs together. This may then introduce a UV sensitivity in the observables.

\subsection{Loop corrections}

We calculate the one-loop Coleman-Weinberg potential during inflation\footnote{\label{postinfl}After inflation, but still in the large field regime $1 < \delta \lesssim 10^4$, the corrections due to the rolling of the field $\dot \phi$ are no longer slow roll suppressed and this might give order one corrections to the RGEs.  Likewise, for the Lorentz violating terms that could be neglected during inflation, see the discussion around (\ref{Lgauge},~\ref{Lfermion}).  Thus in trying to relate the high and low scale physics, one has to deal with threshold corrections as well as the theoretical uncertainty in the running in this post-inflationary field regime. }, which corresponds to very large $\delta > \delta_{\rm end} \sim 10^4$.  We will work in Landau gauge where the ghost fields decouple. We neglect corrections coming from the rolling of the background field, and from the expansion of the universe and from gravity's back reaction, as these are all slow roll suppressed.  We allow for non-minimal fermion and gauge couplings, $\alpha_f$ and $\alpha_A$, which are parameterized as in \eref{alpha}.
The divergent part of the one-loop effective potential is (using dimensional regularization) \cite{CW}
\begin{align}
V_{\rm eff} 
=\frac{\lambda \phi^4}{4} Z_V-\frac{1}{32 \pi^2\eps}
            \[m_h^4 + \sum m_\theta^4 + 3 \sum m_A^4 - 4N_c  m_t^4\] ,
\label{Veff}
\end{align}
with $\epsilon=4-d$, $Z_V =1 +\delta_V$ with $\delta_V$ a counterterm. The first sum in the one-loop contribution is over the three degenerate Goldstone bosons collectively denoted by $\theta$, the second sum over the $W^\pm$ and $Z$ bosons each with 3 polarizations, and we added a factor $4N_c$ for the top quark, a Dirac fermion with $N_c=3$ colors.

All diagrams contributing to the wavefunction normalization of the canonical field $h$ are suppressed by $\O(\delta^{-1})$, and thus to leading order the anomalous dimension $\gamma_h =0$.  For this reason it is useful to rewrite the classical potential in terms of the canonical field, i.e. $\lambda \phi^4 =\tilde\lambda h^{4/3}$, which gives the relation $Z_V = Z_{\tilde \lambda} Z_h^{2/3} = Z_{\tilde \lambda}$. Extraction of the $\delta_V$ counterterm thus gives the full beta function\footnote{Equivalently, $Z_V = Z_\phi^2 Z_\lambda$ from which it follows $\beta_\lambda = \epsilon \delta_V \lambda + 4\lambda \gamma_\phi$.  Since $\gamma_\phi \neq 0$, $\delta_V$ does not give the full $\beta_\lambda$.}
\be
\beta_{\tilde\lambda}=\epsilon\delta_V \tilde \lambda.
\label{betatilde}
\ee  
This can be related directly to the running corrections to $n_s$ and $r$ \eref{observables}.  As we will see below $\beta_{\tilde\lambda} =f(y_t,g_i,\lambda)$ is a function of the top Yukawa coupling, the gauge couplings, and the coupling $\lambda$.  Thus to fully determine the running corrections one needs to calculate the RGEs for  two independent couplings (out of the three $\lambda,\tilde{\lambda}$ and $\mathcal{M}$),  plus the RGEs of the other SM parameters.

\subsubsection{Higgs field}

\label{s:Higgs}

Before tackling the full field content, let us consider only the contributions of the Higgs field to the effective potential.  This simple example explicitly shows EFT renormalizability in the asymptotic regimes, the need for new physics which becomes (at the very least) important in the mid field regime, and that due to these threshold corrections the connection between low and high scales is lost.  We follow the approach put forward for HI in \cite{bezrukov4,shap,critical1} and also in \cite{Enckell:2016xse,Bezrukov:2017dyv}.

We parameterize the potential as $V=\lambda U(h)$ with $h(\phi)$ the canonically normalized Higgs field.  The Higgs contribution to the one loop divergence of the effective action is of the form \eref{Veff} with $m_h^2 = \lambda^2 U''(h)$.  This contribution cannot be absorbed in the counterterm $\delta_V$ over the whole field range; to absorb all divergencies we add the following operator to the original Lagrangian (``h.o." denotes ``higher order"):
\begin{equation}
\mathcal{L}_{\rm{h.o.}}^{(1)}= \lambda^2  c_1  U_1 \equiv  \lambda^2   c_1 [(U''(h))^2-  36 U(h)],
\label{U1}
\end{equation}
where the Wilson coefficient $c_1$ is a new independent parameter. The factor 36 of the 2nd term is chosen such that in the small field regime $U_1 =\O(\delta)$ is suppressed. In this regime $V=(\lambda/4)h^4 +\O(\delta)$ which is EFT renormalizable.  In the large field regime $\delta \gg 1 $ the Higgs mass is suppressed, and $U_1 =-36 \lambda^2  c_1 U +\O(\delta^{-1})$ which is of the same form as the original potential, and it can be absorbed in a shift of the coupling $\lambda\rightarrow \lambda -36\lambda^2 c_1$. This reflects that also the large field regime is EFT renormalizable, as no new counterterms beyond those already present in the original potential are needed.  However, in the midfield regime $\delta \sim 1$, the new operators are not suppressed and they are needed to absorb the divergencies of the non-renormalizable theory.  This is not surprising as $\delta \sim 1$ corresponds to field values of the order of the unitarity bound in the small field regime $\phi \sim \lambda^{-1/2}M \sim \Lambda_{\rm min}$, where new physics is expected.

We have explicitly factored out $\lambda^2$ from the Wilson coefficient in \eref{U1} to give a hierarchical order to the procedure.  The new operator $U_1$ will itself contribute to the one loop corrections at the same order as the two loop corrections coming from the original potential. The new divergencies are absorbed by a new operator $\sim\lambda^3 C_2 U_2$.  In this way, a whole tower of new operators is added, suppressed by increasing powers of the coupling. This allows to consistently truncate the series in the regime where no other order parameters are present. In the two asymptotic regimes this is not needed because $\delta$ respectively $\delta^{-1}$ acts as an order pa\-ra\-me\-ter.  Of course, this procedure, and the hierarchical structure of the operators, represents an assumption on the nature of the UV completion.

We learn a couple of things from this analysis.  First, in the asymptotic regimes the Higgs sector of NHI is EFT renormalizable, but new physics is needed in the mid field regime. Secondly, these threshold corrections break the connection between the low and high scale physics, as the value of $\lambda$ during inflation will depend on the unknown Wilson coefficients. This is both because, as explained above, the coupling in the large field is shifted by the extra $U_1$ contribution, and because the threshold corrections will alter the RG equations in the mid field regime.

\subsubsection{Standard Model spectrum}\label{Sm spectrum}

We will now consider the full SM spectrum.  We will check the d.o.f. entering the one loop diagrams and compare the counterterms needed in the small and large field regime.  If there is a discontinuous jump in d.o.f. and correspondingly in the counterterms, that signals that threshold corrections are needed in the midfield regime.  

The masses of the bosonic fields are given by the covariant expression $(m^2)_a^b = -h^{bc} \del_c \del_a \L$ evaluated on the background, with $h_{ab}={\rm diag}(\gamma,\gamma,k^2)$ the metric on field space, and $\{a,b\}$ running over the Higgs, Goldstone and gauge field. The mass matrix is diagonal, with masses
\begin{align} 
m_h^2 & = \lambda \phi^2 \frac{(3+\delta)}{(1+ \delta)^2} , &
m_\theta^2 & = \lambda \phi^2 \frac{(1+3\delta)}{(1+ \delta)^2 }, &
m_A^2 & = \frac{g^2 \phi^2(1+\delta)}{(1+\alpha_A \delta)} +\frac{(2+n) \alpha_A \delta \lambda \phi^2}{(1+\delta)(1+\alpha_A \delta)},
\end{align}
with $\delta$ evaluated on the background.

In the small field regime the masses approach the Standard Model values. The last term in the gauge boson mass arises from mixing between the Higgs and gauge sector (specifically, because $\Gamma^\phi_{AA}\neq 0$), it is suppressed at large field values for $n < 1$. As will become clear below, its specific form is never important, and we will neglect this term from now on.  We estimate the fermion mass by simply rescaling $q_0 \psi \to \psi$ by a constant factor, to obtain approximately canonically renormalized fermions.  Just as for the gauge field, there will be corrections to this, but these will not be important.  We then get
\be
m_\psi \approx
\frac1{\sqrt{2}}\frac{y}{(1+\alpha_F \delta)} \phi.
\label{yphi}
\ee
It is clear that substituting the masses in the effective potential \eref{Veff} the divergencies cannot be absorbed in field independent counterterms, as the tree-level and one-loop contributions have different field dependence.  The theory is non-renormalizable.  However, an asymptotically renormalizable EFT may be constructed if in these regimes the tree-level and one-loop contributions have at leading order the same field dependence. This happens in the small field regime $\delta \ll 1$ where all masses reduce to their SM masses values, and all divergencies can be absorbed just as in the SM.  Explicitly, 
\begin{align}
V_{\rm eff} |_{\delta \ll 1}
= \frac{\lambda \phi^4}{4}
\[ Z_V -\frac{1}{8 \pi^2\eps} \frac1{\lambda}\(12 \lambda^2 +
  3 \sum g_i^4 - 3 y_t^4
 \) \]+\O(\delta),
\label{Vsmall}
\end{align}
with $g_i = \frac12 \{g^2,g^2,\sqrt{g^2+{g'}^2}\}$, with $g$ and $g'$ the gauge coupling of the SU(2) and U(1) hypercharge gauge groups respectively.

We will check whether also an EFT can be constructed in the large field regime $\delta \gg 1$.  In this limit the effective potential during inflation becomes
\begin{align}
V_{\rm eff}  |_{\rm infl}
= \frac{\lambda\phi^4}{4}
\[ Z_V -\frac{1}{8 \pi^2\eps} \frac1{\lambda}\(
  3 \sum \frac{g^4(1+\delta)^2}{(1+\alpha_A \delta)^2} - N_c \frac{ y_t^4}{(1+\alpha_F \delta)^4} +
 \O(\delta^{-1} )\) \].
\label{Vlarge}
\end{align}
The Higgs field and Goldstone bosons are light during inflation, their contribution to the effective potential is $\O(\delta^{-2})$, and they effectively decouple. Whether the gauge field and fermions decouple or remain in the spectrum depends on their possible non-minimal couplings.  We consider the various cases in turn. 

\begin{itemize}

\item{Case A: $\alpha_A =\alpha_f =0$.}
Consider first minimally coupled gauge bosons and fermions, as in the original NHI model \cite{Germani:2010gm}. The top mass is of the order of the energy scale set by the potential $m_t \sim V^{1/4}$ and is included in the spectrum.  The gauge boson mass, on the other hand, is large, of the order of the cutoff scale \eref{Lambda12} during inflation $m_A^2 = g^2 \Lambda_1^2 /(2\pi^2)$, and should be integrated out.  

Thus, during inflation the Higgs/Goldstone bosons and the gauge fields decouple, and the only d.o.f. in the spectrum is the top quark.  The effective potential is 
\begin{align}\label{CWminimal}
V_{\rm eff} |_{\rm infl}
& = \frac{\lambda\phi^4}{4}\[Z_V -\frac{1}{8 \pi^2\eps}
 \cdot \(- 3 \frac{y_t^4}{\lambda}\) \] +
 \O(\delta^{-1} ) ,
\end{align}
from which we find \eref{betatilde}
\be
\frac{\beta_{\tilde \lambda}}{\tilde \lambda} 
= -\frac{3}{8 \pi^2} \frac{y_t^4}{\lambda}.
\label{betaA}
\ee

Matching to the SM effective field theory at small field values, in the midfield regime the Higgs/Goldstone bosons and the gauge fields are ``integrated back in'', and there are threshold corrections suppressed by the unitarity cutoff $\Lambda_{1,2}$ and $m_A$ respectively.  This new physics is needed to restore unitarity and renormalizability in the UV.

Note, however, that the contribution of the gauge boson to the CW-potential is non-renormalizable.  Consider the case $g \ll 1$ such that there is a hierarchy of scales $V_0^{1/4} \ll m_A \ll \Lambda$.  Then at energy scales $\mu \sim m_A$ the gauge boson is in the spectrum.  To absorb the loop divergence one would need to add a new operator $\L_{\rm h.o} \propto \phi^4 \delta^2$. It is interesting to note that in this particular case the new operator does not generate additional divergences\footnote{In terms of the canonical field in the high field regime $\phi^4\delta^2\approx h^4$.} and (at least at leading order) we do not need to add an infinite tower of higher order operators. On the other hand, unless tuned to be small, this operator will completely change the inflationary dynamics already at tree level.  Thus, to write \eqref{CWminimal} we are implicitly assuming that the physics arising at the cutoff enables us to integrate out the heavy gauge bosons in the usual EFT sense. This scenario can only work with extra assumptions on $\L_{\rm h.o.}$.

\item{Case B:  $\alpha_A= \alpha_{0A}$ constant and $\alpha_f =0$.}

In Case A the gauge fields were heavy during inflation and integrated out. Introducing a non-minimal coupling for the gauge fields $\alpha = \alpha_0$ with $n=0$ in \eref{alpha} will lower their mass and bring them back in the spectrum as now $m_A \sim m_t \sim V^{1/4}$ during inflation, and
\begin{align}
V_{\rm eff} |_{\rm infl}
= \frac{\lambda\phi^4}{4}
\[  Z_V -\frac{1}{8 \pi^2\eps} \frac1{\lambda}\(
  3 \sum_i \frac{g_i^4}{\alpha_{0A} } - 3 y_t^4\) \]  +
 \O(\delta^{-1} ),
\end{align}
from which we find \eref{betatilde}
\be
\frac{\beta_{\tilde \lambda}}{\tilde \lambda} 
= \frac{1}{8 \pi^2} \(  3 \sum_i \frac{g_i^4}{\lambda \alpha_{0A} }-3\frac{y_t^4}{\lambda}\).
\label{betaB}
\ee
In the midfield regime the Higgs/Goldstone bosons are integrated back in, and there are threshold corrections suppressed by the unitarity cutoff $\Lambda_{1,2}$; in addition the gauge contribution is different in the asymptotic regime and additional threshold corrections suppressed by $\Lambda_{A}$ are expected.

Apart from the jump in $\delta_V$ there will also be a jump in the counterterm of the gauge couplings in the midfield regime.  The effect of the non-minimal gauge-gravity coupling is that the effective gauge coupling decreases in the large field regime, and the gauge bosons are coupled more weakly than in case A.  Consider the three-point interaction between SU(2) gauge bosons. The physical gauge coupling for this process is given by $\tilde g = g/(\alpha_0 \delta)$ as follows from \eref{Atilde}.  Then schematically, the amplitude for this process is of the form
\be
\langle A^3 \rangle = Z_{\tilde g} \tilde g + \frac1{8\pi\eps} \tilde g^3
= \frac{ Z_{\tilde g}g  }{\alpha_0 \delta} + \O(\delta^{-3})
\ee
with $Z_{\tilde g} = 1+\delta_{\tilde g} \approx 1$, that is, the running of the gauge couplings in the inflationary regime can be neglected at leading order.

\item{Case C:  $\alpha_i= \alpha_{0i}\delta^{n_i/2}$ with $n_A \geq 1 $ and $n_F \geq 0 $.}

With a large non-minimal coupling for the gauge field, and a non-minimal coupling for the fermions, both gauge fields and fermions are very light during inflation, and decouple from the theory.  To leading order all d.o.f. are weakly coupled and the running can be neglected. The effective potential is now
\begin{align}
V_{\rm eff} |_{\rm infl}
= \frac{\lambda\phi^4}{4}
\[ Z_V+
 \O\(\delta^{-1}, \delta^{-n_A/2}, \delta^{-(1+n_F/2)} \)\],
\end{align}
from which it follows that $\beta_{\tilde \lambda} =0$.
Thus in the inflationary regime the potential is well approximated by its tree level form and we find agreement with the results of \cite{disformal} (where $n_i=1$ was considered).
As in case B, considering the physical gauge $\tilde g$ and Yukawa couplings $\tilde y_t$ of \eref{Atilde}, it follows that loop corrections can be neglected.  There will be threshold corrections suppressed by $\Lambda_{1,2}$, $\Lambda_A$ and $\Lambda_F$.

\end{itemize}

\subsection{Threshold corrections}
\label{s:threshold}

As discussed in the previous subsection, even though in the asymptotic regimes an EFT can be constructed for energies $E \sim V^{1/4} \ll \Lambda_i$ below the unitarity cutoff, threshold corrections are needed to glue the EFTs together in the midfield regime to assure unitarity and renormalizability of the theory.  The unitarity cutoff from interactions in the Higgs sector is always there; we expect that threshold corrections due to integrating back in the Higgs d.o.f. in the mid field regime are suppressed by this scale.  The importance of this type of threshold corrections is smaller for small Higgs coupling, as then the ratio $V^{1/4}/\Lambda_{1,2}$ decreases. This is consistent with the observation that the natural jump in the counterterm $\delta_V$ due to in\-te\-gra\-ting back in the Higgs is proportional to $\lambda^2$, and thus also decreases.  However, these are not the only new scales of physics. In Case A, we expect additional new physics at the scale $m_A$, in Case B at the scale $\Lambda_A$, and in Case C at the scale $\Lambda_A$ and $\Lambda_F$; these scales are unaffected by the size of the Higgs coupling.  

A consistent EFT approach to NHI inflation will require the inclusions of all higher dimensional operators consistent with the theory. As stated before we always work under the  assumption that the UV completion only gives large corrections in the mid field regime, where the operators needed for the consistency of the theory play a significant role. 

The construction discussed in section \ref{s:Higgs} is an example where operators are added to cure the divergencies, which are only important in the mid field regime.  Another approach  followed for HI in \cite{cliffnew,Fumagalli}, is to add only operators suppressed by the field dependent unitarity cutoff of the form (focussing on the most important dimension six operators):
\be
\L_{\rm h.o} = \sum  c_i\frac{\sqrt{V}}{\Lambda^2} O_i + ...
\label{Lho}
\ee
with ellipses denoting the dimension $n\geq 8$ operators, $O_i$ all 4 dimensional SM operators, and $c_i$ the Wilson coefficients that depends on the UV completion.  The new operators can be neglected during inflation as $V^{1/4}_*/\Lambda \ll 1$.   
Different approaches will just give a different parametrization of the uncertainties connected to the mid field regime and consequently will model the running of the parameters connecting the two regimes.

\subsection{Inflationary predictions}

Whatever the explicit parameterization of the UV completion, the net effect will be that the connection between the low and high scales is lost.  The new operators may induce a tree-level shift between the coupling in the small and large field regime (as in the example in section \ref{s:Higgs}), and they will alter the RGE equations in the mid field regime.  The effect is that the couplings get a ``kick'' in the mid field regime. Their values during inflation thus not only depend on the boundary conditions at the electroweak scale, but also on the size of this kick, that is on the new physics parameterized by the unknown Wilson coefficients.\footnote{Moreover, there are theoretical uncertainties in the RGEs as discussed in footnote \eref{postinfl}.}   In this way, the physics during inflation will indirectly depend on the UV physics.  If the inflationary observables depend on these exact values, they inherit the UV dependence, and the model is no longer predictive.

In NHI in all cases A-C discussed, there are threshold corrections and the couplings receive an unknown kick in the mid field regime. Moreover, the linear analysis done in section \ref{rg dependence} shows that the spectral index and tensor-to-scalar ratio \eref{observables} depend on the running of the Higgs coupling.  This dependence is observably large for $\beta_{\tilde \lambda}/(4 {\tilde \lambda}) \sim 1$; since the exact ratio depends on the exact value of the couplings during inflation, which is affected by the kick, the results are UV sensitive if this is the case.  The question thus is whether the running corrections can be of  order one.
In case A and B the beta function  is dominated by the contribution of the top quark $\beta_{\tilde \lambda}/ {\tilde \lambda} \sim 3 y_t^4/(8\pi^2 \lambda) \sim 10^{-3}/\lambda$, where the last estimate  assumes that the couplings $y_t,\lambda$ are not so much different from that in the SM at these energy scales.  We thus expect large running corrections, and large UV dependence, for $\lambda(\mu_*) \lesssim 10^{-3}-10^{-4}$.  Since the Higgs coupling runs to small values at large scales, this would not require excessive tuning. In case C on the other hand all particles are decoupled during inflation, to leading order $\beta_{\tilde \lambda} \approx 0$, and all running corrections are unobservable \cite{disformal}.  In this case, even though the couplings do receive a kick and their value is UV sensitive, the inflationary predictions are robust as they do not depend on the explicit values of the couplings.


\section{Conclusions}\label{conclusions}

In new Higgs inflation (NHI), the Higgs kinetic terms are non-minimally coupled to gravity, enabling the Higgs field to play the role of the inflaton. In this work, we have investigated the renormalization of this model and how this can affect its predictions. In many respects, the situation is analogous to the original Higgs inflation (HI) proposal, where inflation takes place thanks to a non-minimal coupling of the Higgs field to the Ricci scalar. In both models, non-renormalizable interactions are added to the SM and consequently tree level unitarity is violated at energies below the Planck scale.  At every stage in the universe's history the field-dependent unitarity cutoff is above the typical scale set by the Higgs potential, and a renormalizable EFT can be constructed in the asymptotic regimes.  However, the theory is non-renormalizable in the mid field regime, and new operators are needed to absorb the divergencies. This is also the field region where the ratio of potential energy to the unitarity cutoff $V^{1/4}/\Lambda$ is minimized, and we expect the effects of UV physics entering at this new scale to be most important.

To make the above statements more explicit we have calculated the one-loop Coleman-Weinberg potential. In both the small field and large field regime, one can easily identify an order parameter in which a renormalizable EFT can be organized. However, as some fields decouple in the large field regime, the degrees of freedom running in the loop are different in the small field regime, and consequently the counterterms are different.  Threshold corrections are then needed in the mid field regime. We assume that the new physics does not affect the tree level potential significantly.  However, it can still modify the inflationary predictions, as the effect of the new physics in the mid field regime is that it gives a kick to the running couplings.  This kick occurs because the new physics alters the RGE equations in the mid field regime and/or from matching the couplings of the low and high energy theory. The coupling values during inflation thus depend on both the boundary conditions at the electroweak scale and on the unkown UV physics. 

Although details differ, such as the value of the cutoff scale and counterterms, the above qualitative discussion is equally valid for both HI and NHI.  The key difference between the two scenarios is the UV sensitivity of the inflationary observables. In HI, the running corrections to the spectral index and the tensor-to-scalar ratio are suppressed to leading order in the $1/N_\star$-expansion (through a cancellation of different effects \cite{Fumagalli,Fumagalli:2016sof}). For inflation in the universal regime, the predictions are thus robust and do not depend on the UV completion.  This is in contrast with NHI, where the inflationary predictions do depend on the running. This dependence can also be seen as a virtue since in this case the boundary conditions at the EW scale as well as the (for the moment) unknown threshold corrections will leave a direct imprint on the inflationary parameters, which thus can be probed.

The explicit dependencs of $n_s$ and $r$ on the running coupling $\tilde \lambda$ is given in \eref{observables}.  The corrections are of order $\beta_{\tilde \lambda}/\tilde \lambda \sim 10^{-3}/\lambda$,  where the numerical estimate assumes approximately SM running for the top and Higgs self-coupling $\lambda$.  Thus the corrections are important for moderately small couplings $\lambda(\mu_\star) \lesssim 10^{-3}- 10^{-4}$.  However, we have seen that there exists a way to have robust inflationary predictions in NHI as well: with (large) non-minimal couplings for also the gauge and fermion fields all fields decouple in the large field regime and $\beta_{\tilde \lambda} \approx 0$ during inflation.

We believe that focussing on the Coleman-Weinberg corrections, as we have done here, is enough to understand how new physics breaks the connection between low and high field values of the theory, and the effect it has on the robustness of NHI's inflationary predictions. This picture can be made more quantitative by explicitly deriving the full set of RG equations in both asymptotic regimes, which we leave for further work. Another interesting question would be to study the possibility of a ``critical new Higgs inflation" regime, analogous to conventional critical Higgs inflation \cite{Hamada:2014iga,Bezrukov:2014bra,Hamada:2014wna}.


\subsection*{Acknowledgments}

We wish to thank Cristiano Germani, Mikhail Shaposhnikov, Jordy de Vries and Gilberto Tetlalmatzi-Xolocotzi for useful discussions. JF and MP are funded by the Netherlands Foundation for Fundamental
Research of Matter (FOM) and the Netherlands Organisation for Scientific Research (NWO). 

\appendix

 
\section{Renormalization group improved action and in\-fla\-tio\-na\-ry observables}
\label{A:improved}

In this appendix we provide more details on the RG improved action. In particular we check that computing the slow roll parameters and inflationary observables can be done in terms of either the canonical or the non-canonical field, the results are the same. Since this point caused us some worries initially, and since the proof is non-trivial, we think it is worth to derive it explicitly.  Furthermore, once the equivalence of the two procedures has been established, this can be useful to consistently RG improve any model with non-canonical kinetic terms for which is not possible to find an analytic expression for the canonical field.

\subsection{Relations between beta functions and anomalous dimensions}

We start by defining the renormalized fields and couplings in the bare NHI Lagrangian \eqref{L_inf}
\be
\phi_b = Z_\phi^{1/2} \phi, \quad h_b = Z_h^{1/2} h, \quad \lambda_b = Z_\lambda \lambda,\quad \tilde \lambda_b = Z_\lambda \tilde \lambda,
\quad  \M_b = Z_{\M} \M,
\ee
in terms of which this becomes
\be
\L =  -\frac{Z_\phi^3 Z_\lambda}{Z_{\M}^4}  \frac{\lambda \phi^4}{4 \M^4}\frac12 (\partial \phi)^2 - Z_\phi^2 Z_\lambda \frac{\lambda}{4} \phi^4 
= -Z_h \frac12 (\partial h)^2 - Z_h^{2/3} Z_{\tilde \lambda} \frac{\tilde \lambda}{4} h^{4/3}.
\ee
The Lagrangian can be written in terms of either the canonical field $h$ or the field $\phi$.  At tree level $Z_i =1$ and we retrieve the Lagrangians (\ref{L_inf},~\ref{chaL}), with the canonical field defined in \eref{dh}.
For the kinetic terms and the potential to be the same in either language, this gives the relations $Z_h = Z_\phi^3 Z_\lambda{Z_{\M}^{-4}}$ respectively $Z_\phi^2 Z_\lambda = Z_h^{2/3} Z_{\tilde \lambda} $.  Now take the derivative of these relations with respect to the renormalization time  $t = \ln \mu$ to get at one-loop order
\be
\frac{ \beta_\lambda }{\lambda}- 4\frac{\beta_{\M}}{\M} -6 \gamma_\phi +2 \gamma_h =0,
\qquad
\frac{\beta_\lambda }{\lambda}- \frac{\beta_{\tilde \lambda} }{\tilde\lambda}-4 \gamma_\phi + \frac43 \gamma_h =0.
\label{Zrelation1}
\ee
The anomalous dimension is defined as $\gamma_h= -\frac12 \partial_t\log Z_h $ and the beta function is $\partial_t \ln Z_\lambda = -\beta_\lambda/\lambda$, and likewise for the other fields and couplings. Combining the above equations, we find 
\be
\frac{\beta_{\tilde \lambda}}{\tilde \lambda} = \frac13 \frac{\beta_{\lambda}}{\lambda} + \frac83 \frac{\beta_{\M}}{\M},
\label{Zrelation3}
\ee
which could equivalently have been derived from the definition of $\tilde \lambda$ in \eref{lambdat}  (using that $\beta_\lambda = \partial_t \lambda$, and analogously for the other couplings).

%
%

\subsection{RG improved action}

Start with the canonical field. We drop all terms quadratic in and all derivatives of the beta functions and anomalous dimensions, and only keep the leading contribution.  The re\-nor\-ma\-li\-za\-tion group (RG) improved scalar action, i.e. the solution of the Callan-Symanzik equation, is given by
\be
\L = -\frac12 Z_{h}(t) (\partial h)^2 -  \frac{\tilde{\lambda}(t)}{4}  h(t)^{4/3} ,
\label{Lh}
\ee
with
\be\label{relationZZ}
h(t)=e^{-\int_0^t \gamma_h dt'}h,\qquad\,\frac{d\tilde\lambda(t)}{dt}=\beta_{\tilde\lambda},\qquad\,Z_h(t) \approx e^{-2\int_0^t \gamma_h dt'}.
\ee
The potential can be rewritten as $V=\frac14 \tilde \lambda_{\rm{eff}}(t)h^{4/3}$ with $\tilde \lambda_{\rm{eff}}(t)=\tilde \lambda(t)e^{-\frac43 \int_0^t \gamma_h dt'}$.  The canonical field is now defined via $\partial h_c = Z_h^{1/2}(h) \partial h$.  In order to find $h(h_c)$ we approximate $\gamma_h$ as a constant (we work in the same approximation as in section 3, eq. \eqref{approx}), and using the expression of $Z_h(h)$ in \eqref{relationZZ}, we obtain
\be\label{integra}
h_c = \int^{h}_{0} \e^{-\int^{t(h')}_0 \gamma_h \dd t'} \dd h' = \frac{h Z_h^{1/2}(h)}{(1-\gamma_h/3)} 
\quad \Rightarrow \quad
(h Z^{1/2}_h)^{4/3} =\left(1-\frac49 \gamma_h\right) h_c^{4/3}.
\ee
Here we used the explicit form of the normalization scale \eref{mu}, which gives $t(h)=\ln ch^{1/3}$ with $c$ a constant.\footnote{To be precise $t=\ln c(t)h^{1/3}(t)$ but considering the t-dependence inside the log will just give higher order terms in the integral \eqref{integra}.}
Finally, the Lagrangian for the canonical field is
\begin{align}
\L &
= -\frac12 (\partial h_c)^2 -\frac{ \tilde \lambda(h_c)}{4}  h_c^{4/3} \left(1-\frac49 \gamma_h\right).
\label{Lhh}
\end{align}
The story is similar for the non-canonical $\phi$ field. The improved action is
\be
\L_s = -\frac12 \frac{\lambda_{\rm eff} (\phi) \phi^4}{\M^4 (\phi)}Z_\phi(\phi)  (\partial \phi)^2 - \frac{\lambda_{\rm eff} (\phi)}{4} \phi^4 ,
\ee
where now
\be  Z_\phi(\phi)\approx e^{-2\int_0^{t(\phi)}\gamma_\phi dt'},\quad  \lambda_{\rm{eff}}(\phi)=e^{-4\int_0^{t(\phi)}\gamma_\phi dt'}\lambda(\phi),
\ee 
with $\lambda(\phi)$ and $\M(\phi)$ the running couplings,  and $t(\phi)=\ln\phi$ the renormalization time.
We can absorb all dependence on the anomalous dimension in the kinetic term by defining the "canonical" field via
$\phi_c^2 \partial \phi_c = e^{-3\int_0^{t(\phi)}\gamma_\phi dt'} \phi^2\partial \phi$.
Solving as in \eqref{integra} gives $\phi e^{-\int_0^{t(\phi)}\gamma_\phi dt'} = \phi_c (1-\gamma_\phi/3)$. Thus, the improved action becomes
\begin{align}
\L_s
= -\frac12 \frac{\lambda(\phi_c) \phi_c^4}{\M^4(\phi_c)} (\partial \phi_c)^2 - \frac{\lambda(\phi_c)}{4} \phi_c^4 \(1-\frac43 \gamma_\phi\).
\label{Lphi}
\end{align}
%
By comparing the two actions in \eqref{Lhh} and \eqref{Lphi}, it follows that the one loop generalization of \eref{dh} is\footnote{Differentiating the second expression  in \eqref{dhphi} and using \eqref{Zrelation1} returns the first relation in \eqref{dhphi}.}
\be
 \frac{\sqrt{\lambda(t)} \phi_c^2}{2 \M^2(t)} \partial \phi_c = \partial h_c
\quad \Leftrightarrow \quad
\phi_c^3 (1-\gamma_\phi)  = \frac {6\M(t)^2}{\sqrt{\lambda(t)} }  h_c \left(1-\frac13 \gamma_h\right).
\label{dhphi}
\ee

The usual approximation is to drop the $\gamma_i$ correction in the RG improved Lagrangians (\ref{Lhh},~\ref{Lphi}), as the corrections are higher order in the coupling.  That is probably fine for most applications.  In fact, if one calculates the running corrections to $n_s$ and $r$ in either the $h$ or $\phi$ language, one can simply drop these corrections, and the final results are the same (upon using the relation between the beta functions \eref{Zrelation1}). However, if one wants to compare intermediate steps in the calculation in both approaches, for example to show the equivalence of the slow roll parameters, it is essential to keep the $\gamma_i$-corrections in \eref{dhphi}.

\subsection{Slow roll parameters}
Note that $h_c$ correspond to $h$ in the main text where we have omitted the subscript to simplify the notation.

First work in terms of the canonical field, and use the RG improved Lagrangian \eref{Lhh} to calculate (we set $\mpl=1$ in the following)
\be
\frac{V_{h_{c}}}{V} = \frac4{3 h_c} \left(1+ \frac{\beta_{\tilde \lambda}}{4\tilde \lambda} \right)
\ee
where we used $\partial_{h_{c}} \tilde\lambda = \beta_{\tilde\lambda} (\partial t/\partial h_c) \simeq \beta_{\tilde\lambda}/3h_c$.  Now work with $\phi$ and use the Lagrangian \eref{Lphi}.  This gives
\be
\frac{V_{h_{c}}}{V} = \frac{4}{\phi_c} \frac{\partial \phi_c}{\partial h_c}\left(1+ \frac{\beta_\lambda}{4\lambda}\) = \frac{4}{\phi_c^3} \frac{2 \M^2}{\sqrt{\lambda}}\(1+ \frac{\beta_\lambda}{4\lambda}\) =\frac{4}{3h_c} \(1+ \frac{\beta_\lambda}{4\lambda}- \gamma_\phi +\frac13 \gamma_h\) .
\ee
In the first step we used $\partial t /\partial \phi_c= 1/\phi_c$.  In the second step we used the first relation in \eref{dhphi}, and in the last step the last relation in \eref{dhphi}. It is in this last step that it is essential to include the $\gamma_i$-corrections, as now using relation \eref{Zrelation1} it immediately follows that both calculations give the same answer.

The first potential slow parameter is, using the RG improved Lagrangian in terms of $h$ and $\phi$ respectively,
\be
\eps_V = \frac12 \(\frac4{3 h_c}\)^2 \(1+ \frac{\beta_{\tilde \lambda}}{4\tilde \lambda} \)^2
=  \frac12 \( \frac{8}{\phi_c^3} \frac{\M^2}{\sqrt{\lambda}}\)^2  \(1+ \frac{\beta_\lambda}{4\lambda}\)^2.
\ee
In the same way, we find for the second slow roll parameter 
\be
\eta_V =
\frac{4}{9h_c^2} \(1+ 5 \frac{\beta_{\tilde \lambda}}{4\tilde \lambda}\)
=\frac{16 \M^4}{\lambda \phi_c^6} \(1+ \frac34 \frac{\beta_\lambda}{\lambda}+ 2 \frac{\beta_{\M}}{\M}\).
\ee
The two expressions are equal, which can be seen using \eref{dhphi} and \eref{Zrelation1}.

In our approximation computing the number of efolds calculated using the canonical field is trivial:
\be\label{Nh}
N_{\star}=\frac{3}{4}\int dh_c\,h_c \left(1+\frac{\beta_{\tilde \lambda}}{4\tilde\lambda}\right)^{-1}\simeq \frac{3}{8}h^2_{c\star}\left(1-\frac{\beta_{\tilde \lambda}}{4\tilde\lambda}\right)_{\star}.
\ee
It is less trivial working in the $\phi_c$ language, although it should give the same result. Let us check that this is indeed the case.
\be \label{Nphi} 
N_{\star}=\int dh_c \left(\frac{V}{V_{h_{c}}}\right)=\int d\phi_c \left(\frac{\partial h_c}{\partial \phi_c}\right)^2 \frac{\phi_c}{4}\left(1+\frac{\beta_{\lambda}}{4\lambda}\right)^{-1}
\simeq\frac{1}{4^2}\left(1-\frac{\beta_{\lambda}}{4\lambda}\right)_{\star} \int d\phi_c \frac{\lambda(\phi_c)}{\M^4(\phi_c)}\phi_c^5.
\ee
The running couplings in the integrand can be expanded as
\be
\begin{split}
\frac{\lambda(\phi_c)}{\M^4(\phi_c)}&=\frac{\lambda_{\star}}{\M^4_{\star}}+\frac{d(\lambda/\M^4)}{dt}\bigg|_\star\left(\frac{dt}{d\phi_c}\bigg|_\star(\phi_c-\phi_{c\star})+\frac{1}{2!}\frac{d^2t}{d\phi_c^2}\bigg|_\star(\phi_c-\phi_{c\star})^2 +...\right)\\
&=\frac{\lambda_{\star}}{\M^4_{\star}}+\frac{\lambda_{\star}}{\M^4_{\star}}\left(\frac{\beta_{\lambda}}{\lambda}-4\frac{\beta_{\M}}{\M}\right)_{\star}[t(\phi_c)-t(\phi_{c\star})],
\end{split}
\ee
where $t=\ln\phi_c$. Thus \eref{Nphi} becomes
\begin{align}
\label{interm}
N_{\star} &\simeq \frac{\lambda_{\star}}{4^2\cdot6\,\M^4_{\star}}\left(1-\frac{\beta_{\lambda}}{4\lambda}\right)_{\star}\phi^6_{c\star}\left[1-\left(\frac{\beta_{\lambda}}{6\lambda}-\frac{2}{3}\frac{\beta_{\M}}{\M}\right)_{\star}\right] \nn \\
&= 
\frac{\lambda_{\star}}{4^2\cdot6\,\M^4_{\star}}\left(1-\frac{\beta_{\lambda}}{4\lambda}\right)_{\star}
\frac{4\cdot9\M^4_{\star}}{\lambda_{\star}}h^2_{c\star}\left[1+2\left(\gamma_{\phi}-\frac{1}{3}\gamma_h\right)_{\star}\right]\left[1-\left(\frac{\beta_{\lambda}}{6\lambda}-\frac{2}{3}\frac{\beta_{\M}}{\M}\right)_{\star}\right] \nn \\
&=\frac{3}{8}h^2_{c\star} \left(1-\frac{1}{4}\frac{\beta_{\tilde\lambda}}{\tilde\lambda} \right)_{\star}.
\end{align}
In the second step we used \eref{dhphi} to write $\phi_{c\star}$ in terms of the canonical field $h_{c\star}$, and in the last step we used \eqref{Zrelation1} and kept the leading order beta function corrections. This is indeed the same as \eqref{Nh}. Hence, working under the same approximation, both procedures give exactly the same results for both the slow roll parameters and $N_{\star}$. This implies that $n_s$ and $r$ will be also the same computed with the two methods.

\bibliographystyle{utphys}
\bibliography{biblioNHI}

\providecommand{\href}[2]{#2}\begingroup\raggedright\begin{thebibliography}{10}

\bibitem{BasteroGil:2010vq}
M.~Bastero-Gil, A.~Berera, and B.~M. Jackson, ``{Power suppression from
  disparate mass scales in effective scalar field theories of inflation and
  quintessence},'' \href{http://dx.doi.org/10.1088/1475-7516/2011/07/010}{{\em
  JCAP} {\bfseries 1107} (2011) 010},
\href{http://arxiv.org/abs/1003.5636}{{\ttfamily arXiv:1003.5636 [hep-ph]}}.

\bibitem{Linde:1983gd}
A.~D. Linde, ``{Chaotic Inflation},''
\href{http://dx.doi.org/10.1016/0370-2693(83)90837-7}{{\em Phys. Lett.}
  {\bfseries 129B} (1983) 177--181}.

\bibitem{fakir}
R.~Fakir and W.~G. Unruh, ``{Improvement on cosmological chaotic inflation
  through nonminimal coupling},''
\href{http://dx.doi.org/10.1103/PhysRevD.41.1783}{{\em Phys. Rev.} {\bfseries
  D41} (1990) 1783--1791}.

\bibitem{salopek}
D.~S. Salopek, J.~R. Bond, and J.~M. Bardeen, ``{Designing Density Fluctuation
  Spectra in Inflation},''
\href{http://dx.doi.org/10.1103/PhysRevD.40.1753}{{\em Phys. Rev.} {\bfseries
  D40} (1989) 1753}.

\bibitem{bezrukov1}
F.~L. Bezrukov and M.~Shaposhnikov, ``{The Standard Model Higgs boson as the
  inflaton},'' \href{http://dx.doi.org/10.1016/j.physletb.2007.11.072}{{\em
  Phys. Lett.} {\bfseries B659} (2008) 703--706},
\href{http://arxiv.org/abs/0710.3755}{{\ttfamily arXiv:0710.3755 [hep-th]}}.

\bibitem{Germani:2010gm}
C.~Germani and A.~Kehagias, ``{New Model of Inflation with Non-minimal
  Derivative Coupling of Standard Model Higgs Boson to Gravity},''
  \href{http://dx.doi.org/10.1103/PhysRevLett.105.011302}{{\em Phys. Rev.
  Lett.} {\bfseries 105} (2010) 011302},
\href{http://arxiv.org/abs/1003.2635}{{\ttfamily arXiv:1003.2635 [hep-ph]}}.

\bibitem{Kobayashi:2010cm}
T.~Kobayashi, M.~Yamaguchi, and J.~Yokoyama, ``{G-inflation: Inflation driven
  by the Galileon field},''
  \href{http://dx.doi.org/10.1103/PhysRevLett.105.231302}{{\em Phys. Rev.
  Lett.} {\bfseries 105} (2010) 231302},
\href{http://arxiv.org/abs/1008.0603}{{\ttfamily arXiv:1008.0603 [hep-th]}}.

\bibitem{Kamada:2010qe}
K.~Kamada, T.~Kobayashi, M.~Yamaguchi, and J.~Yokoyama, ``{Higgs
  G-inflation},'' \href{http://dx.doi.org/10.1103/PhysRevD.83.083515}{{\em
  Phys. Rev.} {\bfseries D83} (2011) 083515},
\href{http://arxiv.org/abs/1012.4238}{{\ttfamily arXiv:1012.4238
  [astro-ph.CO]}}.

\bibitem{Nakayama:2010sk}
K.~Nakayama and F.~Takahashi, ``{Higgs Chaotic Inflation in Standard Model and
  NMSSM},'' \href{http://dx.doi.org/10.1088/1475-7516/2011/02/010}{{\em JCAP}
  {\bfseries 1102} (2011) 010},
\href{http://arxiv.org/abs/1008.4457}{{\ttfamily arXiv:1008.4457 [hep-ph]}}.

\bibitem{Nakayama:2014koa}
K.~Nakayama and F.~Takahashi, ``{Higgs Chaotic Inflation and the Primordial
  B-mode Polarization Discovered by BICEP2},''
  \href{http://dx.doi.org/10.1016/j.physletb.2014.05.034}{{\em Phys. Lett.}
  {\bfseries B734} (2014) 96--99},
\href{http://arxiv.org/abs/1403.4132}{{\ttfamily arXiv:1403.4132 [hep-ph]}}.

\bibitem{Horndeski:1974wa}
G.~W. Horndeski, ``{Second-order scalar-tensor field equations in a
  four-dimensional space},''
\href{http://dx.doi.org/10.1007/BF01807638}{{\em Int. J. Theor. Phys.}
  {\bfseries 10} (1974) 363--384}.

\bibitem{Kamada:2012se}
K.~Kamada, T.~Kobayashi, T.~Takahashi, M.~Yamaguchi, and J.~Yokoyama,
  ``{Generalized Higgs inflation},''
  \href{http://dx.doi.org/10.1103/PhysRevD.86.023504}{{\em Phys. Rev.}
  {\bfseries D86} (2012) 023504},
\href{http://arxiv.org/abs/1203.4059}{{\ttfamily arXiv:1203.4059 [hep-ph]}}.

\bibitem{ufuk}
U.~Aydemir, M.~M. Anber, and J.~F. Donoghue, ``{Self-healing of unitarity in
  effective field theories and the onset of new physics},''
  \href{http://dx.doi.org/10.1103/PhysRevD.86.014025}{{\em Phys. Rev.}
  {\bfseries D86} (2012) 014025},
\href{http://arxiv.org/abs/1203.5153}{{\ttfamily arXiv:1203.5153 [hep-ph]}}.

\bibitem{cliffnew}
C.~P. Burgess, S.~P. Patil, and M.~Trott, ``{On the Predictiveness of
  Single-Field Inflationary Models},''
  \href{http://dx.doi.org/10.1007/JHEP06(2014)010}{{\em JHEP} {\bfseries 06}
  (2014) 010},
\href{http://arxiv.org/abs/1402.1476}{{\ttfamily arXiv:1402.1476 [hep-ph]}}.

\bibitem{Fumagalli}
J.~Fumagalli and M.~Postma, ``{UV (in)sensitivity of Higgs inflation},''
  \href{http://dx.doi.org/10.1007/JHEP05(2016)049}{{\em JHEP} {\bfseries 05}
  (2016) 049},
\href{http://arxiv.org/abs/1602.07234}{{\ttfamily arXiv:1602.07234 [hep-ph]}}.

\bibitem{Fumagalli:2016sof}
J.~Fumagalli, ``{Renormalization Group independence of Cosmological
  Attractors},'' \href{http://dx.doi.org/10.1016/j.physletb.2017.04.017}{{\em
  Phys. Lett.} {\bfseries B769} (2017) 451--459},
\href{http://arxiv.org/abs/1611.04997}{{\ttfamily arXiv:1611.04997 [hep-th]}}.

\bibitem{bezrukov4}
F.~Bezrukov, A.~Magnin, M.~Shaposhnikov, and S.~Sibiryakov, ``{Higgs inflation:
  consistency and generalisations},''
  \href{http://dx.doi.org/10.1007/JHEP01(2011)016}{{\em JHEP} {\bfseries 01}
  (2011) 016},
\href{http://arxiv.org/abs/1008.5157}{{\ttfamily arXiv:1008.5157 [hep-ph]}}.

\bibitem{Ferrara:2010in}
S.~Ferrara, R.~Kallosh, A.~Linde, A.~Marrani, and A.~Van~Proeyen,
  ``{Superconformal Symmetry, NMSSM, and Inflation},''
  \href{http://dx.doi.org/10.1103/PhysRevD.83.025008}{{\em Phys. Rev.}
  {\bfseries D83} (2011) 025008},
\href{http://arxiv.org/abs/1008.2942}{{\ttfamily arXiv:1008.2942 [hep-th]}}.

\bibitem{Giudice:2010ka}
G.~F. Giudice and H.~M. Lee, ``{Unitarizing Higgs Inflation},''
  \href{http://dx.doi.org/10.1016/j.physletb.2010.10.035}{{\em Phys. Lett.}
  {\bfseries B694} (2011) 294--300},
\href{http://arxiv.org/abs/1010.1417}{{\ttfamily arXiv:1010.1417 [hep-ph]}}.

\bibitem{Barbon:2015fla}
J.~L.~F. Barbon, J.~A. Casas, J.~Elias-Miro, and J.~R. Espinosa, ``{Higgs
  Inflation as a Mirage},''
  \href{http://dx.doi.org/10.1007/JHEP09(2015)027}{{\em JHEP} {\bfseries 09}
  (2015) 027},
\href{http://arxiv.org/abs/1501.02231}{{\ttfamily arXiv:1501.02231 [hep-ph]}}.

\bibitem{George:2015nza}
D.~P. George, S.~Mooij, and M.~Postma, ``{Quantum corrections in Higgs
  inflation: the Standard Model case},''
  \href{http://dx.doi.org/10.1088/1475-7516/2016/04/006}{{\em JCAP} {\bfseries
  1604} no.~04, (2016) 006},
\href{http://arxiv.org/abs/1508.04660}{{\ttfamily arXiv:1508.04660 [hep-th]}}.

\bibitem{shap}
F.~Bezrukov, J.~Rubio, and M.~Shaposhnikov, ``{Living beyond the edge: Higgs
  inflation and vacuum metastability},''
  \href{http://dx.doi.org/10.1103/PhysRevD.92.083512}{{\em Phys. Rev.}
  {\bfseries D92} no.~8, (2015) 083512},
\href{http://arxiv.org/abs/1412.3811}{{\ttfamily arXiv:1412.3811 [hep-ph]}}.

\bibitem{critical1}
F.~Bezrukov and M.~Shaposhnikov, ``{Higgs inflation at the critical point},''
  \href{http://dx.doi.org/10.1016/j.physletb.2014.05.074}{{\em Phys. Lett.}
  {\bfseries B734} (2014) 249--254},
\href{http://arxiv.org/abs/1403.6078}{{\ttfamily arXiv:1403.6078 [hep-ph]}}.

\bibitem{Enckell:2016xse}
V.-M. Enckell, K.~Enqvist, and S.~Nurmi, ``{Observational signatures of Higgs
  inflation},'' \href{http://dx.doi.org/10.1088/1475-7516/2016/07/047}{{\em
  JCAP} {\bfseries 1607} no.~07, (2016) 047},
\href{http://arxiv.org/abs/1603.07572}{{\ttfamily arXiv:1603.07572
  [astro-ph.CO]}}.

\bibitem{Bezrukov:2017dyv}
F.~Bezrukov, M.~Pauly, and J.~Rubio, ``{On the robustness of the primordial
  power spectrum in renormalized Higgs inflation},''
\href{http://arxiv.org/abs/1706.05007}{{\ttfamily arXiv:1706.05007 [hep-ph]}}.

\bibitem{germaniU}
A.~Escrivà and C.~Germani, ``{Beyond dimensional analysis: Higgs and new Higgs
  inflations do not violate unitarity},''
  \href{http://dx.doi.org/10.1103/PhysRevD.95.123526}{{\em Phys. Rev.}
  {\bfseries D95} no.~12, (2017) 123526},
\href{http://arxiv.org/abs/1612.06253}{{\ttfamily arXiv:1612.06253 [hep-ph]}}.

\bibitem{Germani:2011cv}
C.~Germani, ``{Spontaneous localization on a brane via a gravitational
  mechanism},'' \href{http://dx.doi.org/10.1103/PhysRevD.85.055025}{{\em Phys.
  Rev.} {\bfseries D85} (2012) 055025},
\href{http://arxiv.org/abs/1109.3718}{{\ttfamily arXiv:1109.3718 [hep-ph]}}.

\bibitem{disformal}
S.~Di~Vita and C.~Germani, ``{Electroweak vacuum stability and inflation via
  nonminimal derivative couplings to gravity},''
  \href{http://dx.doi.org/10.1103/PhysRevD.93.045005}{{\em Phys. Rev.}
  {\bfseries D93} no.~4, (2016) 045005},
\href{http://arxiv.org/abs/1508.04777}{{\ttfamily arXiv:1508.04777 [hep-ph]}}.

\bibitem{dual1}
A.~B. Balakin and J.~P.~S. Lemos, ``{Non-minimal coupling for the gravitational
  and electromagnetic fields: A General system of equations},''
  \href{http://dx.doi.org/10.1088/0264-9381/22/9/024}{{\em Class. Quant. Grav.}
  {\bfseries 22} (2005) 1867--1880},
\href{http://arxiv.org/abs/gr-qc/0503076}{{\ttfamily arXiv:gr-qc/0503076
  [gr-qc]}}.

\bibitem{dual2}
J.~Beltran~Jimenez, R.~Durrer, L.~Heisenberg, and M.~Thorsrud, ``{Stability of
  Horndeski vector-tensor interactions},''
  \href{http://dx.doi.org/10.1088/1475-7516/2013/10/064}{{\em JCAP} {\bfseries
  1310} (2013) 064},
\href{http://arxiv.org/abs/1308.1867}{{\ttfamily arXiv:1308.1867 [hep-th]}}.

\bibitem{Ema:2015oaa}
Y.~Ema, R.~Jinno, K.~Mukaida, and K.~Nakayama, ``{Particle Production after
  Inflation with Non-minimal Derivative Coupling to Gravity},''
  \href{http://dx.doi.org/10.1088/1475-7516/2015/10/020}{{\em JCAP} {\bfseries
  1510} no.~10, (2015) 020},
\href{http://arxiv.org/abs/1504.07119}{{\ttfamily arXiv:1504.07119 [gr-qc]}}.

\bibitem{Germani:2010ux}
C.~Germani and A.~Kehagias, ``{Cosmological Perturbations in the New Higgs
  Inflation},'' \href{http://dx.doi.org/10.1088/1475-7516/2010/05/019,
  10.1088/1475-7516/2010/06/E01}{{\em JCAP} {\bfseries 1005} (2010) 019},
  \href{http://arxiv.org/abs/1003.4285}{{\ttfamily arXiv:1003.4285
  [astro-ph.CO]}}.
[Erratum: JCAP1006,E01(2010)].

\bibitem{Planck}
{\bfseries Planck} Collaboration, P.~A.~R. Ade {\em et~al.}, ``{Planck 2015
  results. XX. Constraints on inflation},''
\href{http://arxiv.org/abs/1502.02114}{{\ttfamily arXiv:1502.02114
  [astro-ph.CO]}}.

\bibitem{contino}
R.~Contino, ``{The Higgs as a Composite Nambu-Goldstone Boson},''.

\bibitem{kyle}
K.~Allison, ``{Higgs xi-inflation for the 125-126 GeV Higgs: a two-loop
  analysis},'' \href{http://dx.doi.org/10.1007/JHEP02(2014)040}{{\em JHEP}
  {\bfseries 02} (2014) 040},
\href{http://arxiv.org/abs/1306.6931}{{\ttfamily arXiv:1306.6931 [hep-ph]}}.

\bibitem{Hamada:2014iga}
Y.~Hamada, H.~Kawai, K.-y. Oda, and S.~C. Park, ``{Higgs Inflation is Still
  Alive after the Results from BICEP2},''
  \href{http://dx.doi.org/10.1103/PhysRevLett.112.241301}{{\em Phys. Rev.
  Lett.} {\bfseries 112} no.~24, (2014) 241301},
\href{http://arxiv.org/abs/1403.5043}{{\ttfamily arXiv:1403.5043 [hep-ph]}}.

\bibitem{Bezrukov:2014bra}
F.~Bezrukov and M.~Shaposhnikov, ``{Higgs inflation at the critical point},''
  \href{http://dx.doi.org/10.1016/j.physletb.2014.05.074}{{\em Phys. Lett.}
  {\bfseries B734} (2014) 249--254},
\href{http://arxiv.org/abs/1403.6078}{{\ttfamily arXiv:1403.6078 [hep-ph]}}.

\bibitem{Hamada:2014wna}
Y.~Hamada, H.~Kawai, K.-y. Oda, and S.~C. Park, ``{Higgs inflation from
  Standard Model criticality},''
  \href{http://dx.doi.org/10.1103/PhysRevD.91.053008}{{\em Phys. Rev.}
  {\bfseries D91} (2015) 053008},
\href{http://arxiv.org/abs/1408.4864}{{\ttfamily arXiv:1408.4864 [hep-ph]}}.

\bibitem{CW}
S.~R. Coleman and E.~J. Weinberg, ``{Radiative Corrections as the Origin of
  Spontaneous Symmetry Breaking},''
\href{http://dx.doi.org/10.1103/PhysRevD.7.1888}{{\em Phys. Rev.} {\bfseries
  D7} (1973) 1888--1910}.

\end{thebibliography}\endgroup

\end{document}